\documentclass[aps,prx,showpacs,twocolumn,groupedaddress,superscriptaddress]{revtex4-1}

\usepackage{times,xspace}
\usepackage{graphicx,graphics,color,epsfig}
\usepackage{amsmath}
\usepackage{amssymb}
\usepackage{amsfonts}
\usepackage{amsfonts}
\usepackage{epstopdf}
\usepackage{bm}
\usepackage{hyperref}
\usepackage{color}
\usepackage{float}
\usepackage[normalem]{ulem}

\def\be{\begin{equation}}
\def\ee{\end{equation}}
\def\ba{\begin{eqnarray}}
\def\ea{\end{eqnarray}}

\begin{document}

\title{Non-thermal superconductivity in photo-doped multi-orbital Hubbard systems}

\author{Sujay Ray}
%\email{sujay.ray@unifr.ch}
\affiliation{Department of Physics, University of Fribourg, Fribourg-1700, Switzerland}
\author{Yuta Murakami}
\affiliation{Center for Emergent Matter Science, RIKEN, Wako, Saitama 351-0198, Japan}
\author{Philipp Werner}
%\email{philipp.werner@unifr.ch}
\affiliation{Department of Physics, University of Fribourg, Fribourg-1700, Switzerland}

\date{\today}
 
\begin{abstract}
Superconductivity in laser-excited correlated electron systems has attracted considerable interest due to reports of light-induced superconducting-like states. Here we explore the possibility of non-thermal superconducting order in strongly interacting multi-orbital Hubbard systems, using non-equilibrium dynamical mean field theory. We find that a staggered $\eta$-type superconducting phase can be realized on a bipartite lattice in the high photo-doping regime, if the effective temperature of the photo-carriers is sufficiently low. %Moreover, the superconducting order is intertwined with the spin and orbital symmetry which gives rise to a rich phase diagram. Our calculation shows that, 
The $\eta$ superconducting state is stabilized by Hund coupling -- a positive Hund coupling favors orbital-singlet spin-triplet $\eta$ pairing, whereas a negative Hund coupling stabilizes spin-singlet orbital-triplet $\eta$ pairing.
\end{abstract}
\vspace{0.5in}
%\pacs{}
%\keywords{}

\maketitle

{\it{Introduction}}: The study of non-equilibrium quantum systems has become a frontier topic of research in condensed matter physics and quantum field theory \cite{int_qm1,int_qm2,int_qm3,int_qm4}. A particularly interesting topic is the exploration of hidden phases of matter \cite{lat_struc, mang_hid}, that is, long-lived metastable states which can only be accessed via a nonthermal pathway. In strongly correlated electron systems, which are often characterized by competing low-energy states \cite{Dagotto2005}, non-equilibrium protocols offer the intriguing opportunity to manipulate material properties \cite{lat_struc, mang_hid, exciton_insulator,mag_switch, opt_exint, opt_magord, ind_sc_cuprate, phind_sc} and to create novel states with non-thermal electronic orders \cite{nat_rev, AFM_mott, Ono2017, KK_jiajun, yuta_exciton, Kaneko2019, eta_philipp, Tindall2019, oneband_jiajun, Werner2021,Murakami2023,Takashi2002,Rosch2008,Murakami2022}. A promising route for realizing such non-equilibrium states is laser driving. This allows to create Floquet states \cite{Oka2009,Wang2013} or photo-doped states \cite{Iwai2003}, which under suitable circumstances can favor electron pairing and superconductivity. % which gave rise to the exciting field of light induced superconductivity. 
In photo-doped large-gap Mott insulators, the charge excitations survive for a long time \cite{Sensarma2010,Eckstein2011,Strohmaier2010,Lenarcic2014}. Rather than relaxing back to a (heated) equilibrium state, the charge excitations thermalize within the Hubbard bands, resulting in the formation of a metastable non-thermal state 
%After the intra-band relaxation, %the charge distribution becomes well-defined and 
%the non-thermal state 
which 
can be characterized by effective temperatures of the charge carriers. % in the different Hubbard bands.

A broad range of theoretical  tools has been employed to simulate photo-induced non-equilibrium states \cite{Maeshima2005,Golez2012,Shinjo2017,Sandri2013,Eckstein2013,Tacogne2018, Rincon2018, Zawadzki2019,Schueler2020,Bittner2021,Perfetto2022}. 
%\textcolor{red}{[check if all these DMRG works are about photo-induced states]}
One of the challenges in laser-driven systems is the heating effect, which tends to suppress the emergence of electronic orders. Sophisticated protocols have been developed to realize effectively cold metastable  systems. Entropy cooling techniques \cite{entropy_cooling,eta_philipp} have proven useful in overcoming relaxation bottlenecks and enabled the preparation of effectively cold non-equilibrium states of photo-doped Mott systems. Alternatively, such states can be realized by the coupling to suitable fermionic \cite{NESS_martin} or bosonic baths \cite{Eckstein2013}. %Entropy cooling techniques and steady-state approaches with a coupling to electron baths 
Both techniques 
have been successfully used to simulate nonthermal superconducting (SC) states in the single band repulsive Hubbard model \cite{eta_philipp,oneband_jiajun,tri_chiral}. SC states with staggered order parameter can be stabilized over a wide photo-doping region on bipartite lattices \cite{oneband_jiajun},  while chiral SC order with a $120^\circ$ phase twist has been induced on a triangular lattice \cite{tri_chiral}. Here, we extend these studies to multi-orbital Hubbard systems, more specifically the two-orbital model with strong repulsive interactions. %It is, however, extremely challenging to study real-time dynamics in multi-orbital system in a similar manner. The process becomes numerically too heavy to drive the system for long enough time and obtain a cold system.

Nonthermal orders in the two-orbital repulsive Hubbard model have been previously studied. These investigations include nonthermal spin and orbital orders in the three-quarter filled (Kugel-Khomskii) model \cite{KK_jiajun,Kugel_Khomskii,Kugel_Khomskii_1}, and excitonic orders in the model with crystal field splitting \cite{yuta_exciton}. A photo-induced enhancement of a spin-triplet pairing susceptibility has also been reported \cite{enhanced_pairing}, but due to computational limitations, this study could not reach a cold enough state for symmetry breaking, and it did not consider the staggered order that we will identify below as the dominant nonthermal phase. % has not been achieved so far to indicate a superconducting state in two-orbital system.

In this letter, we employ the recently developed non-equilibrium steady state (NESS) technique \cite{NESS_martin} to explore the photo-doped two-orbital repulsive Hubbard model. With this method we are able to achieve low enough effective temperatures to demonstrate the emergence of 
nonthermal SC states which are specific to multi-orbital systems. 
%staggered $\eta$ pairing superconductivity in the heavily photo-doped half-filled system. Because of the active spin and orbital degrees of freedom, different types of pairing are possible. %, depending on the spin and orbital symmetry. 
%Hund coupling plays an important role in determining the symmetry of the dominant superconducting state, and we may expect a rich nonequilibrium phase diagram as a function of the Hubbard repulsion $U$, Hund coupling $J$, photo-doping concentration and effective temperature $\beta_\text{eff}$. %The rest of the letter is arranged as follows - first, we describe our model and non-equilibrium steady state (NESS) DMFT method to realize a photo-doped system. Next, we show the results of our calculations and describe the optical conductivity, which can be compared to experiments. And finally, we conclude with the outlook and our future direction.
%\\

%\label{sec_2}
{\it{Model and Method}}: We consider the two-orbital Hubbard model with Hamiltonian
\begin{align}
&H = -t_\text{hop}\sum_{\left<ij\right>,\sigma}\sum_{\alpha=1,2} c^{\dagger}_{i,\alpha\sigma}c_{j,\alpha\sigma} \nonumber \\
&\hspace{0mm} + U\sum_{i}\sum_{\alpha=1,2}n_{i,\alpha\uparrow}n_{i,\alpha\downarrow} - \mu \sum_{i}\sum_{\alpha=1,2}\left(n_{i,\alpha\uparrow} + n_{i,\alpha\downarrow} \right) \nonumber \\
&\hspace{0mm} + (U-2J)\sum_{i,\sigma}n_{i,1\sigma}n_{i,2\Bar{\sigma}} %\nonumber \\
%&& 
+ (U-3J)\sum_{i,\sigma}n_{i,1\sigma}n_{i,2\sigma} 
\label{eq_1}
\end{align}
on a bipartite lattice. Here, $\alpha$ and $\sigma$ denote the orbital and spin indices respectively, $t_\text{hop}$ is the nearest neighbor hopping amplitude between sites $i$ and $j$, $U$ is the intra-orbital Hubbard repulsion, $J$ the Hund coupling, and $\mu$ the chemical potential. The equilibrium ground state of the half-filled model at large $U$ is a Mott insulator with antiferro spin ($J>0$) or orbital ($J<0$) order. % {\red {(for $J<0$ it is antiferro orbital order. Should we mention or clarify that??)}} %with intertwined spin and orbital order, which can be understood with Kugel-Khomskii model\cite{Kugel_Khomskii, Kugel_Khomskii_1}. 
% [KK is for the quarter-filled system?]
Away from half-filling, for $J>0$, orbital-singlet spin-triplet superconductivity appears over a wide range of dopings between quarter filling ($n=1/4$) and three-quarter filling ($n=3/4$) \cite{Hoshino, Hoshino_2, enhanced_pairing}, and it is interesting to ask if a nonthermal manifestation of this order exists. 
%At quarter filling, the system is described by the Kugel-Khomskii model and exhibits a nontrivial interplay of spin and orbital order. In the presence of a crystal field splitting, spin-orbital (excitonic) order is realized near half-filling \cite{Hoshino}. It is interesting to investigate the role of these intertwined orders in non-equilibrium systems, and to search for the appearance of novel types of electronic orders in photo-doped Mott states. Here, we focus on the  emergence of superconducting orders. % and analyze how superconductivity is influenced by the spin-orbital order. 

\begin{figure}[t]
\includegraphics[width=0.48\textwidth]{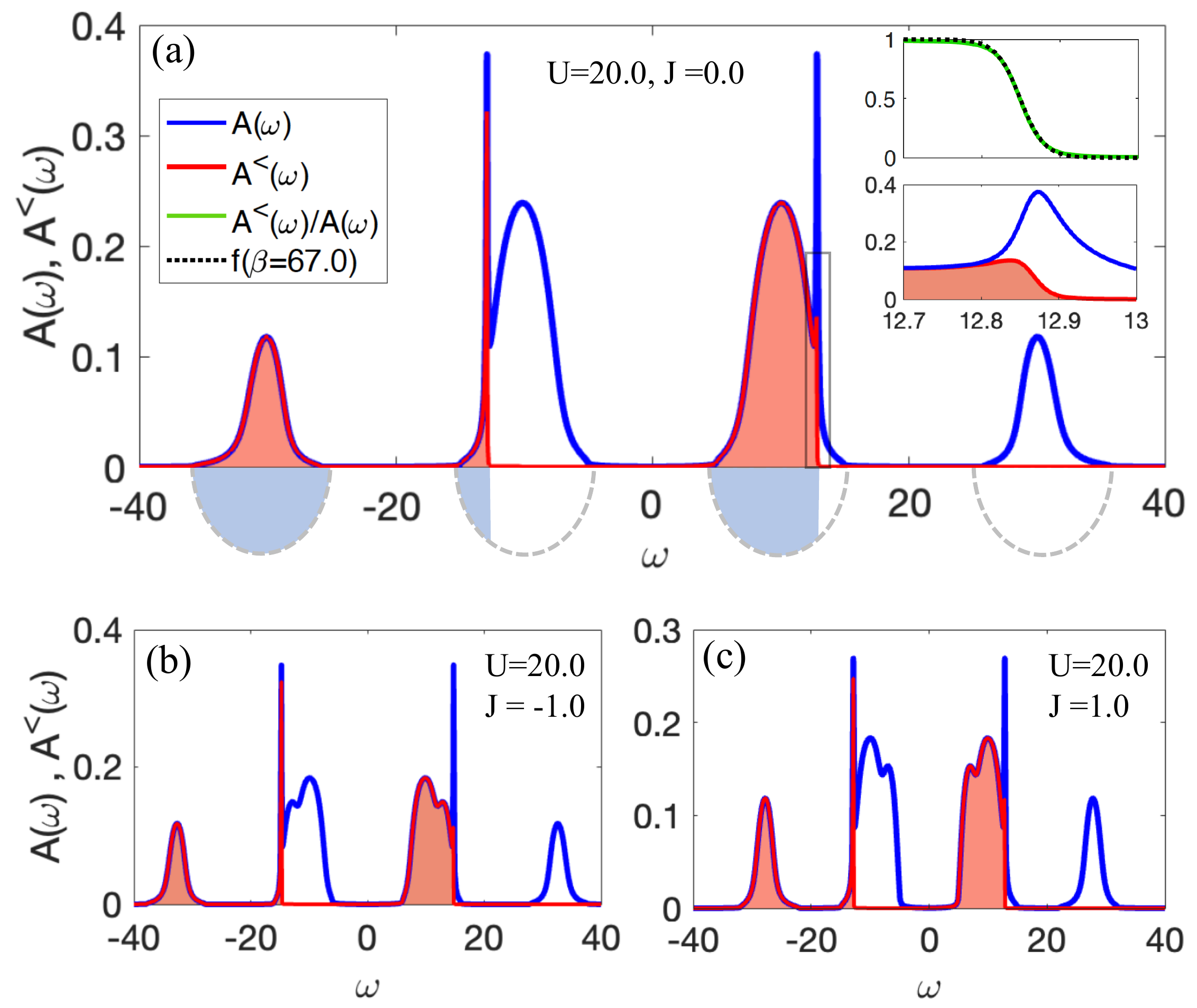}
\caption{Spectra of the photo-doped two-orbital model with (a) $U=20$, $J=0$, (b) $U=20$, $J=-1$, and (c) $U=20$, $J=1$. The occupied DOS is indicated by the filled area. The four fermion baths and their occupation are indicated below the spectra in (a) by the gray dashed lines and filled blue areas, respectively. The insets in panel (a) show a zoom of the region indicated by the gray box and a fit to the Fermi distribution function with $\beta=67.0$ (black dashed line). 
%This Fermi function reproduces $A^{<}(\omega)/A(\omega)$ (green line) very well, which indicates that the triplons have a well-defined effective temperature.
}
\label{fig1}
\end{figure}

To create an effectively cold photo-doped state, we employ the NESS formalism \cite{NESS_martin, oneband_jiajun}, where the system is weakly coupled to cold fermion baths at each site. By choosing the densities of states (DOS) $\rho_b(\omega)$ and chemical potentials $\mu_b$ of these baths appropriately, we can inject (remove) electrons into (from) the upper (lower) Hubbard band, and stabilize a photo-doped state. Here, we use fermion baths with a semi-elliptic DOS %$\rho_{fb}(\omega)$ 
with a coupling strength $\Gamma$. The hybridization function $D_b(\omega)$ of these baths is given by
%\begin{eqnarray}
$D_b(\omega)=$ $\Gamma\rho_{b}(\omega)=\Gamma \sqrt{W_{b}^{2}-(\omega - \omega_{b})^{2}}$,
%\end{eqnarray}
where $W_b$ denotes the half-bandwidth and $\omega_b$ indicates the center of the energy spectrum of the bath. 
%The purpose of this fermion bath is to inject electrons in the upper Mott-Hubbard band and take out electrons from the lower Mott-Hubbard band and thus creating a photodoped non-equilibrium state. By carefully tuning the parameter $\omega_{b}$, we can shift the fermion baths exactly to  the upper and lower Mott-Hubbard bands which facilitates the doping process. We can also control the amount of doping by including a suitable chemical potential $\mu_b$ for the fermion baths.
%We use in total four baths: two baths are coupled to the upper and lower Hubbard bands to create a large population of triply occupied sites (triplons) and singly occupied sites (singlons). The other two baths are attached to the high energy spectral features associated with fully occupied and empty sites, to prevent a nonthermal occupation of these states. 
We use four fermion baths with $\omega_{b}=\pm U/2, \pm 3U/2$. The two baths with $\omega_{b}=\pm U/2$ are chosen to coincide with the lower and upper Hubbard bands, to create a large population of triply occupied sites (triplons) and singly occupied sites (singlons). The other two baths, a completely empty one with $\omega_{b}= 3U/2$ and a completely filled one at $\omega_{b}= -3U/2$, are added to prevent the production of high-energy fully occupied (quadruplon) and empty (holon) states, see Fig.~\ref{fig1}(a). %For two-orbital model, 
%The photo-dopped state we are interested in consists of triplons and singlons, with an admixture of doublons. Quadruplon and holon states are separated by an energy $U$ from lower and upper Mott-Hubbard bands and should not be excited. However, since we want to create strongly photo-dopped states with an almost complete population inversion, a number of quadruplons and holons would be produced in the NESS approach in the absence of those second baths. 

The NESS solution of the two-orbital Hubbard model is obtained on an infinitely connected Bethe lattice using non-equilibrium Dynamical Mean-Field theory (DMFT) \cite{Aoki2014} and a non-crossing approximation (NCA) impurity solver \cite{Eckstein2010}. To study uniform or staggered SC orders, we use the Nambu Keldysh formalism to write the DMFT self-consistency condition as $\Delta(t,t^{\prime})=t_{0}^{2}\gamma G(t,t^{\prime})\gamma + D(t,t^{\prime})$. Here, $t_0=W/4$ is the quarter-bandwidth of the Bethe-lattice DOS, which is used as the unit of energy. The hybridization function $\Delta$, impurity Green's function $G$ and hybridization function of the  fermion baths $D=\sum_b D_b$ can be represented by matrices in the Nambu basis, see Supplemental Material (SM), while the choice of the $\gamma$ matrix determines whether the solution corresponds to staggered or uniform pairing \cite{eta_philipp}.

In the calculations, we use $U=20$, $J=\pm 1$, $\Gamma=0.016$-$0.032$ and $W_b=5$. A typical spectral function $A(\omega)$ for a photo-doped system with triplon density $=0.474$ is shown in Fig.~\ref{fig1}(a). The occupied density of states $A^{<}(\omega)$ is indicated by the filled area. An effective temperature $T_\text{eff}$ of the system can be defined by comparing the ratio $A^{<}(\omega)/A(\omega)$ to a (shifted) Fermi distribution function $f_\beta(\omega)$ for inverse temperature $\beta$. The inset of Fig.~\ref{fig1}(a) shows that $A^{<}(\omega)/A(\omega)$ can be very well fitted with $f_{\beta=67.0}(\omega)$, which indicates that the triplons (and singlons) in the non-equilibrium state have a well-defined and low effective temperature. We also see sharp quasi-particle peaks at the edges of the Hubbard bands, which %quasi-particle peaks %are exclusive features of non-equilibrium photo-doped states, which consist of mostly triplons and singlons. These peaks become sharper with lower temperature indicating the stability of the non-equilibrium state.
are associated with the remaining doublon states. 
In equilibrium, at low temperature, a single doublon state dominates and the energy separation between the upper and lower Hubbard bands is roughly $U+J$ ($J>0$) or $U-5J$ ($J<0$).  
In a photo-doped system, all doublon states contribute to the spectrum, which exhibits peaks with a separation between $U-5J$ and $U+J$. 
%Hence for positive (negative) $J$, states inside (above) the gap are populated (Figs.~\ref{fig1}(b,c)). 
Hence for $J>0$ ($J<0$), the separation between the Hubbard bands decreases (increases), see Figs.~\ref{fig1}(b,c) and SM. 
%For a positive $J$, the energy gap decreases because the spin aligned doublon states have an effective repulsion $U-3J$, and analogously the gap increases for negative $J$ (see Figs.~\ref{fig1}(b) and (c)). 
%\textcolor{red}{[should be opposite?]}
A large enough gap ensures a slow relaxation across the gap \cite{Sensarma2010,Eckstein2011} and is necessary for the stability of the photo-doped state, which sets limits to values of $J$. %In our calculations, we are able to obtain stable spectra with $J=1$ for $U=20$ as shown in Fig.~\ref{fig1}(b) and (c).\\

{\it{Results}}:  In the two-orbital model we can define six types of SC orders, depending on the spin and orbital symmetry. The three orbital-singlet spin-triplet orders are measured by \cite{Hoshino}
\begin{eqnarray}
p_{i,s\nu} &=& \frac{1}{4}\sum_{\alpha\sigma\alpha^{\prime}\sigma^{\prime}}c^{\dagger}_{i,\alpha\sigma}\epsilon_{\alpha\alpha^{\prime}}(\sigma^{\nu}\epsilon)_{\sigma\sigma^{\prime}}c^{\dagger}_{i,\alpha^{\prime}\sigma^{\prime}},
\label{eq_3}
\end{eqnarray}
and the three spin-singlet orbital-triplet orders by 
\begin{eqnarray}
p_{i,o\nu} &=& \frac{1}{4}\sum_{\alpha\sigma\alpha^{\prime}\sigma^{\prime}}c^{\dagger}_{i,\alpha\sigma}(\sigma^{\nu}\epsilon)_{\alpha\alpha^{\prime}}\epsilon_{\sigma\sigma^{\prime}}c^{\dagger}_{i,\alpha^{\prime}\sigma^{\prime}},
\label{eq_4}
\end{eqnarray}
where $\sigma^{\nu}$ denotes the three Pauli matrices with $\nu=x,y,z$ and the antisymmetric tensor is defined as $\epsilon=i\sigma^{y}$. In analogy to antiferromagnetic order in magnetically ordered systems, a staggered SC order ($\eta$ order) can be defined, where $  p_{i,\alpha} (\alpha=sx,sy,sz,ox,oy,oz)$ has opposite signs on the two sublattices $A$ and $B$, such that 
\begin{eqnarray}
\eta^{+}_{i,\alpha}=\delta_i^{A/B}p_{i,\alpha},
\label{eq_5}
\end{eqnarray}
with $\delta_i^{A(B)}=1(-1)$ for $i$ on the $A(B)$ sublattice. 

%In the two-orbital Hubbard Hamiltonian without 
Without 
Hund coupling, the local Hamiltonian is fourfold degenerate (spin degeneracy and orbital degeneracy) and at half-filling, for sufficiently large $U$, each site is doubly occupied in the ground state. The photo-excitation creates singlon-triplon pairs. In the extreme photo-doping limit, all doubly occupied sites are converted to singlons and triplons, and if the fluctuations between these states become large, we may expect the emergence of superconductivity with order parameters given by Eqs.~\eqref{eq_3} or \eqref{eq_4}, similar to the $\eta$ pairing states in the photo-doped single-band Hubbard model \cite{eta_philipp,oneband_jiajun,Murakami2023,Rosch2008}.
An important distinction however is that 
%However, there is one important distinction. Unlike in single band model where excited states are spinless doublons and holons, 
the triplons and singlons in the two-orbital model have spin and orbital degrees of freedom. %Thus, spin and orbital orders may also appear and compete with the SC states. 
The Hund coupling $J$ lifts the degeneracy between the SC orders in the half-filled system and favors a spin ($J>0$) or orbital ($J<0$) moment at each site. It also stabilizes the SC pairing, so that an $\eta$ SC order can be realized at sufficiently large photo-doping. % which is shown by considering an effective model for photo-doped system.

\begin{figure}[t]
\includegraphics[width=0.45\textwidth]{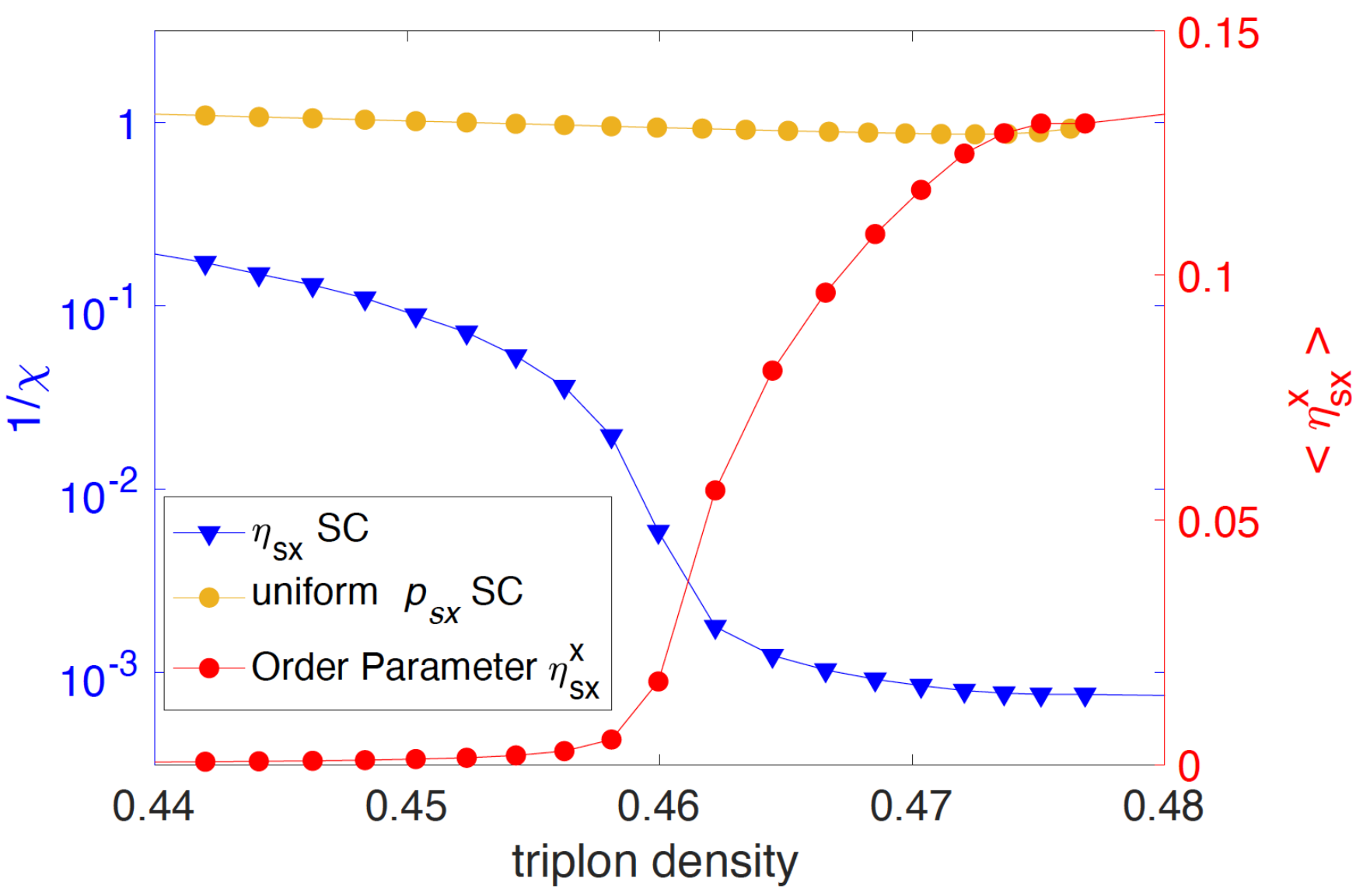}
\caption{SC order parameter $\eta^{x}_{sx}$ and inverse susceptibility $1/\chi_{SC}$ as a function of the triplon density for $U=20$, $J=1$, $\Gamma=0.02$ and $P_\text{seed}=10^{-4}$. While $\chi_{SC}$ for $\eta$ pairing is as high as $10^{3}$ for large triplon densities, for uniform pairing it remains of order $1$.
}
\label{fig2}
\end{figure}

{\it{ Effective model}}: Physical properties of photo-doped states can be well captured by an effective model derived from the Schrieffer-Wolff transformation (SWT) or 2nd order perturbation theory~\cite{oneband_jiajun}. For $J=0$ and half filling, the effective model for complete doping, where only tripons and singlons exist, can be derived by projecting the  transformed Hamiltonian onto the singlon and triplon subspace (see SM) \cite{SWT}, %which yields
\begin{eqnarray}
H_\text{eff} &=& H_{t_{\text{hop}}^2/U}^{st} + H_{t_{\text{hop}}^2/U}^{tt} + H_{t_{\text{hop}}^2/U}^{ss},
\end{eqnarray}
where the low-energy effective interaction between triplons and singlons is divided into three terms representing the singlon-triplon, triplon-triplon and singlon-singlon interaction. Each term can be expressed in terms of spin, orbital and $\eta$-spin operators as 
\begin{eqnarray}
\label{eq_7}
H_{t_{\text{hop}}^2/U}^{st} &=& -\frac{4t_\text{hop}^2}{U} \sum_{\left<ij\right>,\alpha} \left( \eta_{i\alpha}^{+}\eta_{j\alpha}^{-} + \eta_{i\alpha}^{-}\eta_{j\alpha}^{+} \right) \nonumber \\
&& - \frac{4t_\text{hop}^2}{3U} \sum_{\left<ij\right>} \left( \frac{1}{2} + 2\boldsymbol{s_i} . \boldsymbol{s_j} \right) \left( \frac{1}{2} + 2\boldsymbol{\tau_i} . \boldsymbol{\tau_j} \right), \\
%H_{t^2/U}^{ss} &=& H_{t^2/U}^{tt} = 
H_{t_{\text{hop}}^2/U}^{ss(tt)} &=& 
 \frac{2t_\text{hop}^2}{U} \sum_{\left<ij\right>} \left( \frac{1}{2} + 2\boldsymbol{s_i} . \boldsymbol{s_j} \right) \left( \frac{1}{2} + 2\boldsymbol{\tau_i} . \boldsymbol{\tau_j} \right),
\label{eq_8}
\end{eqnarray}
where $\eta^{+}_{i\alpha}$ is defined in Eq.~\eqref{eq_5}, $\eta^{-}_{i\alpha}$ is the Hermitian conjugate $(\eta^{+}_{i\alpha})^{\dagger}$, $\boldsymbol{s}_i=\sum_{\alpha}c^{\dagger}_{i,\alpha\sigma}\boldsymbol{\sigma}_{\sigma\sigma^{\prime}}c_{i,\alpha\sigma^{\prime}}$ the spin moment, and $\boldsymbol{\tau}_i=\sum_{\sigma}c^{\dagger}_{i,\alpha\sigma}\boldsymbol{\sigma}_{\alpha\alpha^{\prime}}c_{i,\alpha^{\prime}\sigma}$ the orbital moment. As can be seen from Eqs.~\eqref{eq_7} and \eqref{eq_8}, in addition to $\eta$-spin interaction terms, there are also spin interactions, orbital interactions and spin-orbital composite terms. %\red{\sout{and we may expect associated long-range orders.}} 
The temperature scales for all types of $\eta$ orders are set by the coefficients which are $\sim t_\text{hop}^{2}/U$. Nonzero Hund coupling lifts the degeneracy between these $\eta$ orders, because it affects the energies of the doublon states involved in the second order hopping processes. The  exchange coupling for the $\eta$ pairing channels $\alpha$ (the first term in Eq.~\eqref{eq_7}) is modified as $t_\text{hop}^2/U \rightarrow t_\text{hop}^2/\tilde{U}_\alpha$, where $\tilde{U}_\alpha$ is a function of $U$ and $J$.  For $J > 0$, $\tilde{U}_{sx}$ takes the smallest value
and favors spin-triplet $\eta_{sx}$ superconducting order (see SM). 

\begin{figure}[t]
\includegraphics[width=0.48\textwidth]{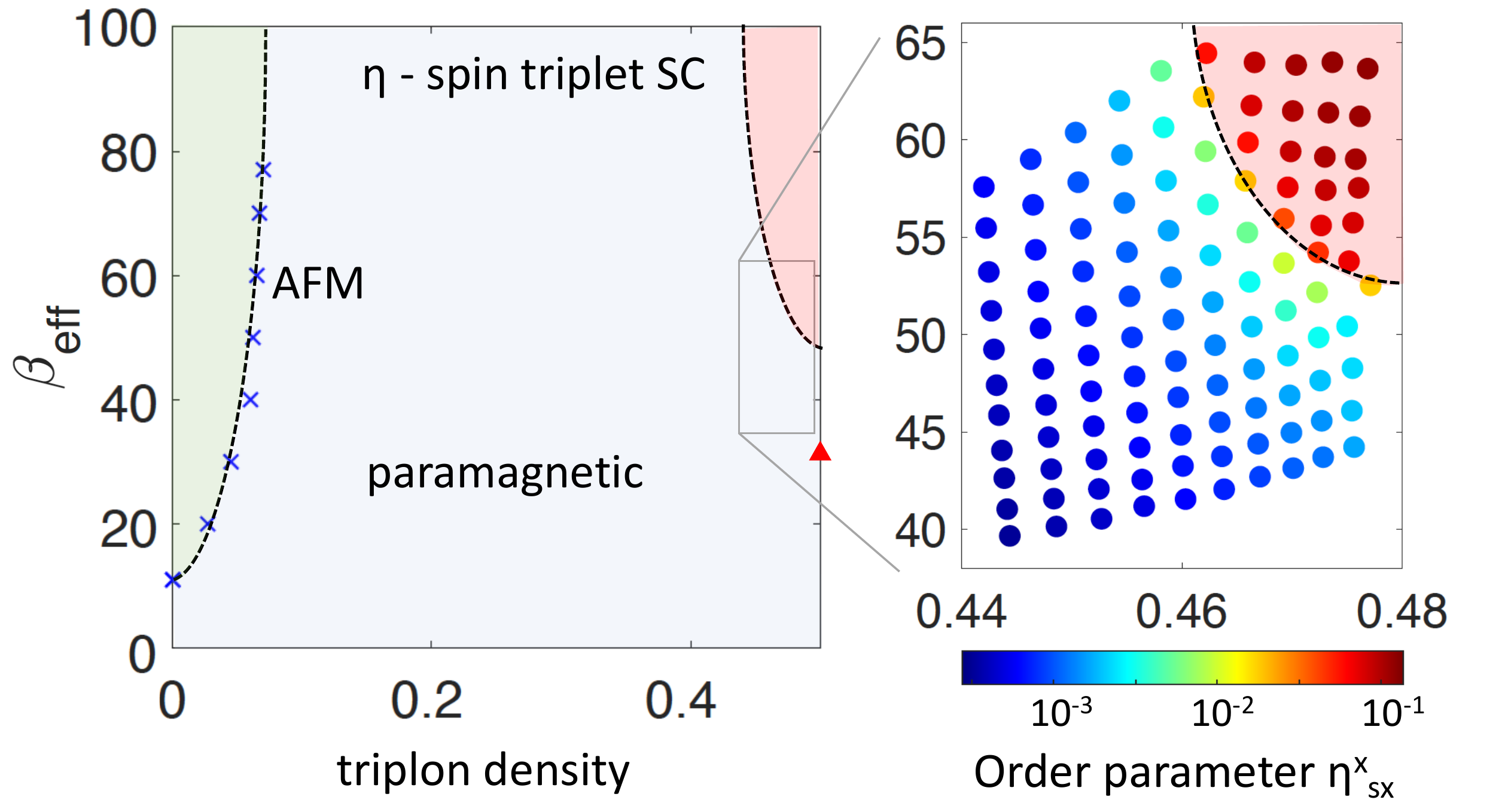}
\caption{Phase diagram of the two-orbital model with $U=20$, $J=1$ and $\Gamma=0.02$ in the space of $\beta_\text{eff}$ and triplon density. The AFM phase appears in the undoped and weakly photo-doped region, while the $\eta$ SC state is stabilized in the high photo-doping region. The red triangle marks the MF transition temperature at triplon density $0.5$. The right panel shows the color map of the order parameter Re[$\eta^{+}_{sx}$] in the high photo-doping region. 
}
\label{fig3}
\end{figure}

A mean-field (MF) decoupling may be applied to the effective model to obtain a MF Hamiltonian (see SM). This MF calculation suggests that an $\eta$ SC state can be achieved at low $T_\text{eff}$, and that the Hund coupling enhances spin-triplet or orbital-triplet superconductivity, depending on $J$. However, compared to the non-equilibrium DMFT results, MF overestimates the SC transition temperature (red triangle in Fig.~\ref{fig3}). This is because MF neglects temporal fluctuations. Moreover, the effective model is valid only in the large-$U$ limit and the DMFT solution is affected by the bath couplings.

%{\it{$\eta$-superconducting states}}: 
{\it{DMFT results}}: 
To detect a SC state, we apply a small seed field $P_\text{seed}$ and couple it to the SC order we are interested in. For example, if we look for the $p_{sx}$ SC order, we add the term
%\begin{eqnarray}
$H_\text{pair}=P_\text{seed}\sum_i (p_{i,sx} + h.c.)$ %\nonumber \\
%&=& P_\text{seed}\sum_i(c^{\dagger}_{i1\uparrow}c^{\dagger}_{i2\uparrow} - c^{\dagger}_{i1\downarrow}c^{i\dagger}_{2\downarrow} + h.c.)
%\end{eqnarray}
to the Hamiltonian \eqref{eq_1} and measure the order parameter $\left< (p_{sx} + h.c.) \right> $. For a $\eta$ SC state, we use a staggered local seed field with opposite signs on the two sublattices and measure the SC order parameter $\left< \eta^{x}_{\alpha} \right>$ and SC susceptibility $\chi_{SC}$ as
\begin{align}
&\left< \eta^{x}_{\alpha} \right>=\text{Re}[\left< \eta^{+}_{\alpha} \right>]=\frac{\left< \eta^{+}_{\alpha} + \eta^{-}_{\alpha} \right>}{2}, %\\
% \chi_{SC}&=&\frac{\left< \eta^{x}_{\alpha} \right>}{P_\text{seed}}.
 \quad  \chi_{SC}=\frac{\left< \eta^{x}_{\alpha} \right>}{P_\text{seed}}.
 \end{align}

The effective temperature and band filling can be controlled by tuning the temperature and chemical potentials of the fermion baths, which allows us to map out the phase diagram of the non-equilibrium photo-doped state in the space of triplon density and $T_\text{eff}$. For $J=0$, all $\eta$ pairing phases are degenerate and the highest transition temperature is $T_{\text{eff}} = 1/62.6$. A nonzero positive (negative) Hund coupling favors the spin triplet (orbital triplet) $\eta$ phases, and the corresponding transition temperatures increase.

Figure~\ref{fig2} plots the SC susceptibility $\chi_{SC}$ and order parameter $\left<\eta^{x}_{sx}\right>$ as a function of the triplon density for $U=20$ and $J=1$. %As expected from the effective model, the 
The 
$\eta$ SC state appears at high triplon density ($\gtrsim 0.46$), as indicated by the large increase in $\chi_{SC}$. We observe an order parameter of the order of $10^{-1}$, which for $P_\text{seed}=10^{-4}$ corresponds to $\chi_{SC} \sim 10^{3}$. Importantly, this order parameter is almost unchanged if the seed is reduced to $P_\text{seed}=10^{-5}$, which demonstrates that the order is due to a spontaneous symmetry breaking (see SM).
%\textcolor{red}{[you can try to add the results for OP and the inverse chi for Pseed=1e-5 in the figure. If it looks too busy, we show this in the SM. By the way, is the transition first order? Maybe try an even smaller seed?]}
For comparison, we also plot in Fig.~\ref{fig2} the susceptibility for the uniform order $p_{sx}$, which remains small.

The $\beta_\text{eff}$ vs triplon density phase diagram for the same interaction parameters is shown in Fig.~\ref{fig3}. At half filling, without any photo-doping, we have an antiferromagnetic (AFM) state, which is quickly suppressed with increasing triplon density. Over a wide region of intermediate photo-dopings, the system is in a normal paramagnetic state. As we approach the extreme doping limit, an $\eta$ spin-triplet SC state is realized below a sufficiently low effective temperature ($\beta_\text{eff} \gtrsim 52.0$).

\begin{figure}[t]
\includegraphics[width=0.4\textwidth]{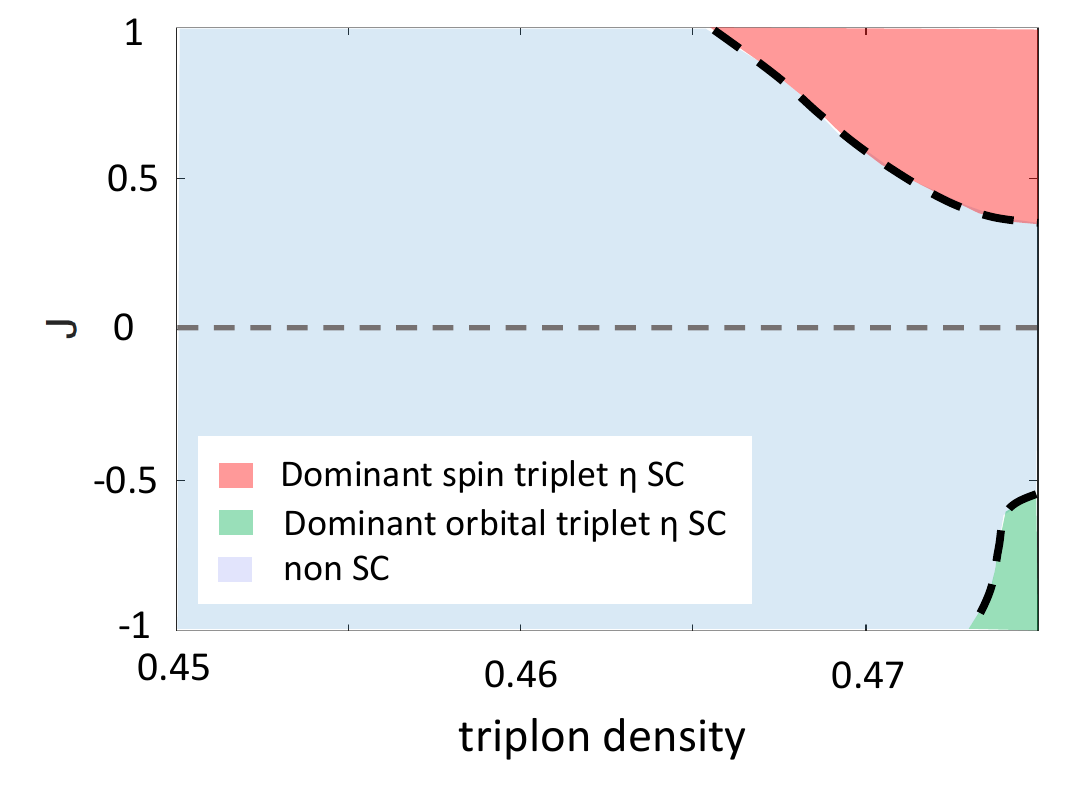}
\caption{Nonquilibrium phase diagram in the space of triplon density and Hund coupling $J$, for $U=20$ and $\beta_\text{eff}=60$.
}
\label{fig4}
\end{figure}

As mentioned earlier, in the two-orbital system there can be various SC orders with different spin and orbital symmetries. Thus, in Fig.~\ref{fig4}, we also show the phase diagram for fixed $U=20$ and $\beta_\text{eff}=60$ in the space spanned by the triplon density and Hund coupling. % (the effective inverse temperature is in the range $\beta_\text{eff}=60\pm1.5$). 
%\textcolor{red}{[since we need to cut 1/2 page, I suggest we only show and discuss pancel (c) in the main text, while (a,b) can be moved to the SM]}
The data for the orbital-triplet order $\eta^{x}_{ox}$ and spin-triplet order $\eta^{x}_{sx}$, on which this phase diagram is based, are shown in the SM. Increasing $|J|$ favors the $\eta$ SC order in both cases with a dominant $\eta^{x}_{sx}$ pairing for positive $J$. %For $J>0.3$ both $\eta^{x}_{ox}$ and $\eta^{x}_{sx}$ appear. However, since the order parameter for $\eta^{x}_{sx}$ is larger than for $\eta^{x}_{ox}$, we can conclude that the SC phase has a dominant spin-triplet $\eta^{x}_{sx}$ order. We note that with decreasing $J$, the SC phase is pushed towards higher doping levels and even lower temperatures would be required to drive the system into SC state. 
%\textcolor{red}{[actually, how do you deal with the effective temperature in these plots? do diferent points correspond to different temperatures, or do you keep the effective temperature fixed?- The phase diagram is obtained by keeping the bath temperature and the coupling to the bath fixed. This should give roughly the same temperature for all the states. However it is very difficult to maintain the system at exactly the same temperature for different doping. So, the temperature of the systems is $\beta_{eff}\sim 60$ with tolerance of $\pm 1.5$]}
Insights into the $J<0$ case can be obtained by noting that the $J>0$ model can be mapped (qualitatively) to the $J<0$ model by %flipping the spin and orbital degrees of freedom as 
the transformation 
$c_{i,1\downarrow} \rightarrow c_{i,2\uparrow}$ and $c_{i,2\uparrow} \rightarrow c_{i,1\downarrow}$ \cite{Hoshino_2}. This exchanges the roles of spin and orbital and it is thus natural that
$\eta^{x}_{ox}$ pairing becomes dominant for negative $J$. The smaller region of $\eta^{x}_{ox}$ pairing compared to $\eta^{x}_{sx}$ is due to the fact that the relevant interaction $\tilde U_\alpha$ is larger for $J<0$ than for $J>0$, which reduces the critical temperature ($\propto t_\text{hop}^2/\tilde{U}_\alpha$) for $J<0$. 

{\it {Optical conductivity}}: We finally calculate the optical conductivity and study how the optical response changes as we approach the SC phase from the normal phase. The steady state optical conductivity $\sigma(\omega)$ is calculated for the Bethe lattice from the local Green's function, following Ref.~\cite{oneband_jiajun}.

\begin{figure}[t]
\includegraphics[width=0.48\textwidth]{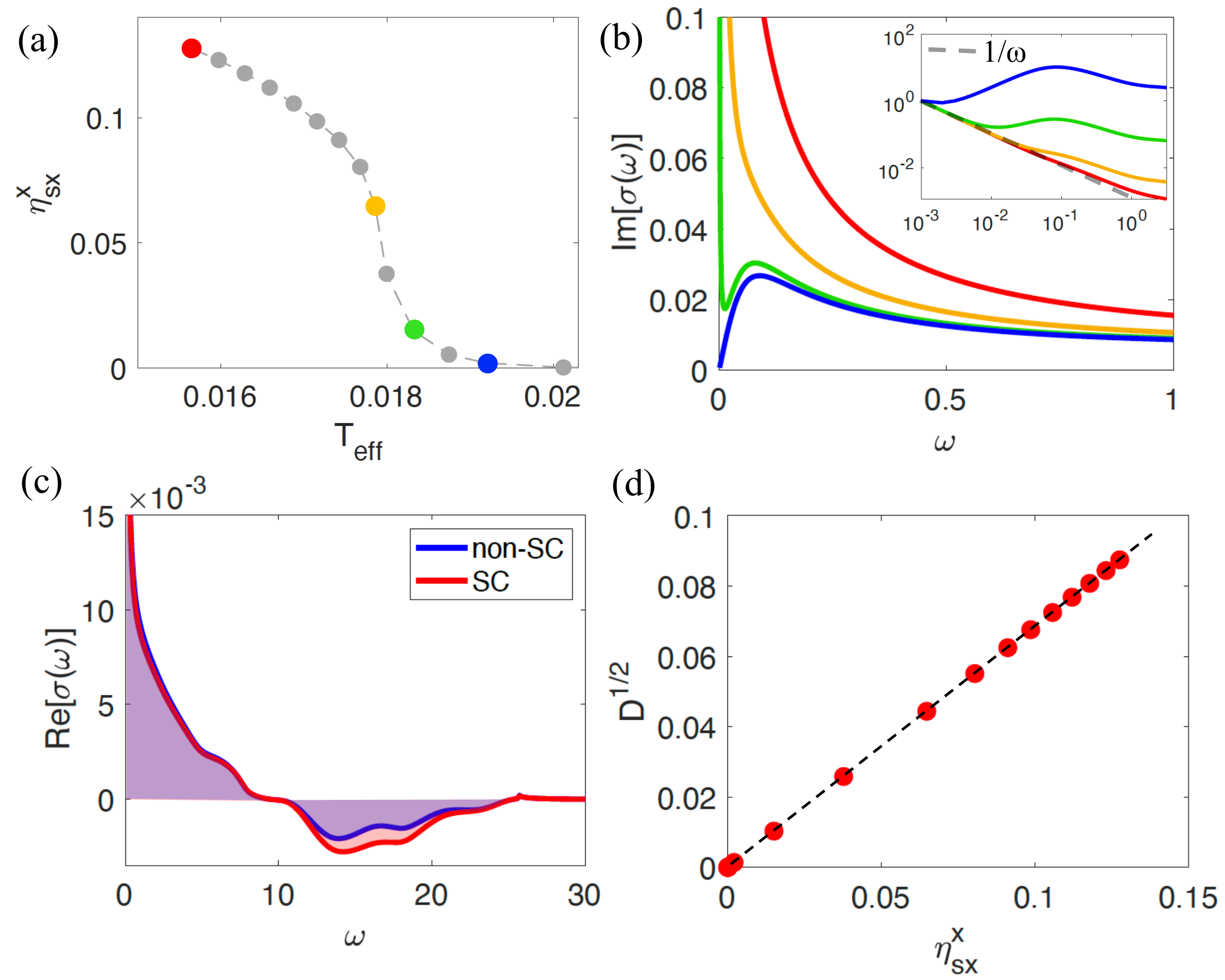}
\caption{Optical conductivity results for $U=20$, $J=1$ and for triplon density 0.473. (a) Order parameter $\eta^{x}_{sx}$ as a function of effective temperature $T_\text{eff}$, (b) imaginary part and (c) real part of the optical conductivity as a function of $\omega$. The different colors in (b) and (c) correspond to the corresponding data points in (a). The inset of (b) shows the log-log plot of (shifted) Im$[\sigma(\omega)]$. (d) Square root of the SC phase stiffness $D$ as a function of the order parameter $\eta^{x}_{sx}$.    
}
\label{fig5}
\end{figure}

Figure~\ref{fig5}(a) shows the order parameter $\eta^{x}_{sx}$ as a fuction of $T_\text{eff}$ for the triplon density $0.473$, $U=20$ and $J=1$. The SC transition temperature is $T_c \sim 0.019$. In both the normal and SC state, in addition to a narrow Drude peak, which we associate with the doublon population, the real part of $\sigma(\omega)$ shows a range of negative spectral weight around $\omega \sim 15$ (panel (c)). The latter comes from triplon-singlon recombination processes in the population inverted non-equilibrium system, which are associated with local energy changes in the range from $U-5J$ to $U+J$. 
%\textcolor{red}{[should it be $U+J$ here?]}
The imaginary part, however, shows a distinct low-frequency behavior in the SC and non-SC states (panel (b)). In the non-SC state (blue), Im$[\sigma(\omega)]$ decreases to $0$ as $\omega \rightarrow 0$. The onset of superconductivity (green dot in panel (a)) is marked by a diverging Im[$\sigma(\omega)$] as $\omega\rightarrow 0$, as shown by the green curve in panel (b). All the states which have a lower $T_\text{eff}$ than this state, and hence are superconducting, exhibit a $1/\omega$ divergence at low $\omega$. The inset of Fig.~\ref{fig5}(b) shows the log-log plot of appropriately shifted Im[$\sigma(\omega)$] and we can clearly observe the $1/\omega$ behavior below $\omega\approx 10^{-1}$ for the SC states (red and  yellow). As $T_\text{eff}$ increases, the energy region over which the $1/\omega$ scaling can be observed decreases and eventually disappears in the non-SC state (blue). We note that all the curves in Fig.~\ref{fig5}(b) show an approximate power-law decay for $\omega > 10^{-1}$. However, the normal state behavior is different from a $1/\omega$ decay, as can be seen from the inset. The $1/\omega$ scaling at low frequencies thus provides evidence for the realization of a nonthermal SC state. 

The $1/\omega$ divergence of Im[$\sigma(\omega)$] implies that we can define the SC phase stiffness $D$ from the relation $$\lim\limits_{\omega \to 0} \text{Im}[\sigma(\omega)]=\frac{D}{\omega}.$$ In Fig.~\ref{fig5}(d) we plot $\sqrt{D}$ as a function of the order parameter $\eta^{x}_{sx}$ for fixed $U=20$, $J=1$ and observe a linear scaling. This $D\propto |\eta|^{2}/U$ scaling is reminiscent of the two-fluid model and holds in the entire SC region \cite{oneband_jiajun}. 

{\it {Conclusions}}: We demonstrated that upon photo-doping, staggered $\eta$ SC states can be realized in a small region near triplon density $0.5$ in the half-filled two-orbital repulsive Hubbard model.  Unlike in the single band model, the relevant local states have spin and orbital degrees of freedom, which potentially allows to realize different SC states. We showed that a positive Hund coupling enhances orbital-singlet spin-triplet $\eta$ SC pairing whereas a negative Hund coupling favors orbital-triplet spin-singlet $\eta$ pairing. Due to the large $U$ needed to stabilize the non-equilibrium states, the highest $T_\text{eff}$ for realizing multi-orbital $\eta$ pairing states are about an order of magnitue lower than in the single band case \cite{oneband_jiajun}. However, in the realistic case where the energy unit is $\sim$0.5 eV, the SC state can still be realized up to high effective temperatures of about 100~K. It will be interesting to explore similar $\eta$ pairing states in other photo-doped multi-orbital systems and to check if SC states coexisting with spin or orbital orders can be realized. %{\red {Any proposed experiment or realistic model calculation for multi-orbital $\eta$ pairing.}}  
%A possible material to search for these novel $\eta$ pairing states is NiO, which is often described by a half-filled two-orbital Hubbard model. \red{The gap in NiO may however be too small to support a long-lived almost completely population inverted state, and the coupling to other bands or phonons may further destabilize this state.}

{\it Acknowledgements} We thank J. Li and M. Eckstein for providing their NESS DMFT solver and J. Li and S. Hoshino for helpful discussions. The calculations have been run on the beo05 and beo06 clusters at the University of Fribourg. We acknowledge support from ERC Consolidator Grant No. 724103 and from SNSF Grant No. 200021-196966.

%\clearpage

%\appendix
%
%\section{Effective inter-layer magnetic field}\label{Sec:AppA}
%
%\begin{tabular}{| p{3.5cm} | p{2.5cm} | p{1.5cm} |}
%    \hline
%     Transitions & ${\bf r}_d$ & $J_{\perp}/J_{D}$ \\ \hline \hline
%    Phase I to phase II & ${\bf R}_1/2$ & -0.57 \\ \hline
%    Phase II to phase III & ${\bf R}_2/2$ & -0.4 \\ \hline
%    Phase III to phase IV & $({\bf R}_1+{\bf R}_2)/2$ & 0.3  \\ \hline
%    Phase IV to phase V & ${\bf 0}$ & 1.6 \\
%    \hline
%\end{tabular}
%\\
%\\
%
%\begin{tabular}{| p{2.8cm} | p{2.4cm} | p{2.4cm} |}
%    \hline
%     quantities & $J_D=0.53$ & $J_D=1.0$ \\ \hline \hline
%    $q_1$ & -0.23 & -0.23 \\ \hline
%    $q_2$ & 0.22 & 0.22 \\ \hline
%    $q_3$ & -0.25 & -0.24 \\ \hline
%    $q_4$ & 0.22 & 0.24 \\ \hline
%    $q_5$ & -0.20 & -0.22 \\ \hline
%    $(x_1,y_1)$ in umit of moir\'e lattice vector R=35.5 & (0.20 , 0.07) & (0.20 , 0.07) \\ \hline
%    $(x_2,-y_2)$ in umit of moir\'e lattice vector R=35.5 & (0.35 , -0.09) & (0.35 , -0.09) \\ \hline
%    ${\bf p}_1$ & 0.033$\hat{\bf x}$ + 0.035$\hat{\bf y}$ & 0.033$\hat{\bf x}$ + 0.035$\hat{\bf y}$ \\ \hline
%    ${\bf p}_2$ & -0.033$\hat{\bf x}$ - 0.035$\hat{\bf y}$ & -0.034$\hat{\bf x}$ - 0.037$\hat{\bf y}$ \\ \hline
%    $Q_{xx}$ & -0.035 & -0.037 \\ \hline
%    $Q_{yy}$ & -0.001 & -0.001 \\ \hline
%    $Q_{xy}$ & 0.021 & 0.020 \\
%    \hline
%     \end{tabular}


\begin{thebibliography}{99}

\bibitem{int_qm1} D. N. Basov,  R. D. Averitt, and D. Hsieh , Nat. Mat. {\bf 16}, 1077 (2017).

\bibitem{int_qm2} C. Giannetti, M. Capone, D. Fausti, M. Fabrizio, F. Parmigiani, and D. Mihailovic, Advances in Physics {\bf 65}, 58-238 (2016).

\bibitem{int_qm3} T. Oka and S. Kitamura, Annual Review of Condensed  Matter Physics {\bf 10}, 387 (2019).

\bibitem{int_qm4} A. dela Torre, D. M. Kennes, M. Claassen, S. Gerber, J. W. McIver, and M. A. Sentef, Rev. Mod. Phys. {\bf 93}, 041002 (2021). 

\bibitem{lat_struc} H. Ichikawa, S. Nozawa, T. Sato, A. Tomita, K. Ichiyanagi, M. Chollet, L. Guerin, N. Dean, A. Cavalleri, S.-i. Adachi, {\it et al.}, Nat. Mater. {\bf 10}, 101 (2011).

\bibitem{mang_hid} L. Stojchevska, I. Vaskivskyi, T. Mertelj, P. Kusar, D. Svetin, S. Brazovskii, and D. Mihailovic, Science {\bf 344}, 177 (2014).

%\bibitem{ind_sc_cuprate} D. Fausti, R. I. Tobey, N. Dean, S. Kaiser, A. Dienst, M. C. Hoffmann, S. Pyon, T. Takayama, H. Takagi, and A. Cavalleri, Science {\bf 331}, 189 (2011).

\bibitem{exciton_insulator} S. Mor, M. Herzog, D. Golez, P. Werner, M. Eckstein, N. Katayama, M. Nohara, H. Takagi, T. Mizokawa, C. Monney, and J. Stahler, Phys. Rev. Lett. {\bf 119}, 086401 (2017).

\bibitem{Dagotto2005} E. Dagotto, Science 309, 257 (2005). 

\bibitem{mag_switch} T. Li, A. Patz, L. Mouchliadis, J. Yan, T. A. Lograsso, I. E. Perakis, and J. Wang, Nature {\bf 496}, 69 (2013).

\bibitem{opt_exint} R. Mikhaylovskiy, E. Hendry, A. Secchi, J. H. Mentink, M. Eckstein, A. Wu, R. Pisarev, V. Kruglyak, M. Katsnel-son, T. Rasing, et al., Nat. Commun. {\bf 6}, 8190 (2015).

\bibitem{opt_magord} A. Kirilyuk, A. V. Kimel, and T. Rasing, Rev. Mod. Phys. {\bf 82}, 2731 (2010).

\bibitem{ind_sc_cuprate} D. Fausti, R. I. Tobey, N. Dean, S. Kaiser, A. Dienst, M. C. Hoffmann, S. Pyon, T. Takayama, H. Takagi, and A. Cavalleri, Science {\bf 331}, 189 (2011).

\bibitem{phind_sc} M. Mitrano, A. Cantaluppi, D. Nicoletti, S. Kaiser, A. Perucchi, S. Lupi, P. Di Pietro, D. Pontiroli, M. Ricco, S. R. Clark, et al., Nature {\bf 530}, 461 (2016).


\bibitem{nat_rev} D. Basov, R. Averitt, and D. Hsieh, Nat. Mater. {\bf 16}, 1077 (2017).

\bibitem{AFM_mott} P. Werner, N. Tsuji, and M. Eckstein, Phys. Rev. B {\bf 86}, 205101 (2012).

%\bibitem{AFM_relax} D. Golez, J. Bonca, M. Mierzejewski, and L. Vidmar, Phys. Rev. B {\bf 89}, 165118 (2014).

\bibitem{Ono2017} A. Ono and S. Ishihara, Phys. Rev. Lett. {\bf 119}, 207202 (2017).

\bibitem{KK_jiajun} J. Li, H. U. R. Strand, P. Werner and M. Eckstein , Nat. Com. {\bf 9}, 4581 (2018).

\bibitem{yuta_exciton} P. Werner, and Y. Murakami, Phys. Rev. B {\bf 102}, 241103(R) (2020).

%\bibitem{AFM_screening} D. Golez, L. Boehnke, H. U. R. Strand, M. Eckstein, and P. Werner, Phys. Rev. Lett. {\bf 118}, 246402 (2017).

\bibitem{Kaneko2019} T. Kaneko, T. Shirakawa, S. Sorella, and S. Yunoki. Phys. Rev. Lett. {\bf 122}, 077002 (2019).

\bibitem{eta_philipp} P. Werner, J. Li, D. Golez, and M. Eckstein, Phys. Rev. B {\bf 100}, 155130 (2019).

\bibitem{Tindall2019} J. Tindall, B. Buca, J. R. Coulthard, and D. Jaksch, Phys. Rev. Lett. {\bf 123}, 030603 (2019).

\bibitem{oneband_jiajun} J. Li, D. Golez, P. Werner, and M. Eckstein, Phys. Rev. B {\bf 102}, 165136 (2020).

\bibitem{Werner2021} P. Werner and Y. Murakami, Phys. Rev. B {\bf 104}, L201101 (2021).

\bibitem{Murakami2023} Y. Murakami, S. Takayoshi, T. Kaneko, A. M. L\"auchli, and P. Werner, Phys. Rev. Lett. {\bf 130}, 106501 (2023).

\bibitem{Takashi2002}  A. Takahashi, H. Gomi, and M. Aihara, Phys. Rev. B {\bf 66}, 115103 (2002). 

\bibitem{Rosch2008} A. Rosch, D. Rasch, B. Binz, and M. Vojta, Phys. Rev. Lett. {\bf 101}, 265301 (2008). 

\bibitem{Murakami2022} Y. Murakami, S. Takayoshi, T. Kaneko, Z. Sun, D. Golež, A. J. Millis, and P. Werner, Communications Physics {\bf 5}, 23 (2022). 

\bibitem{Oka2009} T. Oka and H. Aoki. Phys. Rev. B {\bf 79}, 081406(R) (2009).

\bibitem{Wang2013} Y. H. Wang, H. Steinberg, P. Jarillo-Herrero, and N. Gedik, Science {\bf 342}, 453 (2013). 

\bibitem{Iwai2003} S. Iwai, M. Ono, A. Maeda, H. Matsuzaki, H. Kishida, H. Okamoto, and Y. Tokura, Phys. Rev. Lett. {\bf 91}, 057401 (2003).

\bibitem{Sensarma2010} R. Sensarma, D. Pekker, E. Altman, E. Demler, N. Strohmaier, D. Greif, R. J\"ordens, L. Tarruell, H. Moritz, and T. Esslinger, Phys. Rev. B {\bf 82}, 224302 (2010).

\bibitem{Eckstein2011} M. Eckstein and P. Werner, Phys. Rev. B {\bf 84}, 035122 (2011).

\bibitem{Strohmaier2010} N. Strohmaier, D. Greif, R. Jordens, L. Tarruell, H. Moritz, T. Esslinger, R. Sensarma, D. Pekker, E. Alt- man, and E. Demler, Phys. Rev. Lett. {\bf 104}, 080401 (2010).

\bibitem{Lenarcic2014} Z. Lenarcic and P. Prelovsek, Phys. Rev. B {\bf 90}, 235136 (2014). 

\bibitem{Maeshima2005} N. Maeshima, and K. Yonemitsu, J. Phys. Soc. Jpn. {\bf 74}, 2671 (2005).

\bibitem{Golez2012} D. Golez, J. Bonca, L. Vidmar, and S. A. Trugman, Phys. Rev. Lett. {\bf 109}, 236402 (2012).

\bibitem{Shinjo2017} K. Shinjo and T. Tohyama, Phys. Rev. B {\bf 96}, 195141 (2017).

\bibitem{Sandri2013} M. Sandri and M. Fabrizio, Phys. Rev. B {\bf 88}, 165113 (2013).

\bibitem{Eckstein2013} M. Eckstein and P. Werner, Phys. Rev. Lett. {\bf 110}, 126401 (2013).

%\bibitem{noneq_DMRG1} J. Eisert, M. van den Worm, S.R. Manmana, and M. Kastner, PRL {\bf {111}}, 260401 (2013).

%\bibitem{noneq_DMRG2} N. Schlünzen, J.-P. Joost, F. Heidrich-Meisner, and M. Bonitz, Phys. Rev. B {\bf {95}}, 165139 (2017).

%\bibitem{noneq_DMRG3} J. Chem. Phys. {\bf {148}} J. S. Kretchmer and G. K.-L. Chan, J. Chem. Phys. {\bf {148}}, 054108 (2018).

%\bibitem{noneq_DMRG4} N. Schlünzen, S. Hermanns, M. Bonitz, and C. Verdozzi, Phys. Rev. B {\bf {93}}, 035107 (2016).

%\bibitem{noneq_ED} E. Arrigoni, M. Knap, and W. von der Linden, Phys. Rev. Lett. {\bf {110}}, 086403 (2013).

\bibitem{Tacogne2018} N. Tancogne-Dejean, M. A. Sentef, and A. Rubio, Phys. Rev. Lett. {\bf 121}, 097402 (2018).

\bibitem{Rincon2018} J. Rincon, E. Dagotto, A. E. Feiguin, Phys. Rev. B {\bf 97}, 235104 (2018).

\bibitem{Zawadzki2019} K. Zawadzki and A. E. Feiguin, Phys. Rev. B {\bf 100}, 195124 (2019).

\bibitem{Schueler2020} M. Sch\"uler, U. De Giovannini, H. H\"ubener, A. Rubio, M. A. Sentef, T. P. Devereaux, and P. Werner, Phys. Rev. X {\bf 10}, 041013 (2020).

\bibitem{Bittner2021}  N. Bittner, D. Golez, M. Casula, and P. Werner, Phys. Rev. B {\bf 104}, 115138 (2021).

\bibitem{Perfetto2022} E. Perfetto, Y. Pavlyukh, and G. Stefanucci, Phys. Rev. Lett. {\bf 128}, 016801 (2022).

%\bibitem{noneq_ten} J. Thoenniss, A. Lerose, and D. A. Abanin, arXiv:2205.04995v1.



\bibitem{entropy_cooling} P. Werner, M. Eckstein, M. Müller, and G. Refael, Nat. Comm. {\bf 10}, 5556 (2019).

%\bibitem{Li2021} Jiajun Li and Martin Eckstein. Phys. Rev. B {\f 103}, 045133 (2021).
\bibitem{NESS_martin} J. Li and M. Eckstein, Phys. Rev. B {\bf 103}, 045133 (2021).


\bibitem{tri_chiral} J. Li, M. M\"uller, A. J. Kim, A. L\"aeuchli, and P. Werner, Phys. Rev. B {\bf 107}, 205115 (2023).


\bibitem{Kugel_Khomskii} K. I. Kugel and D. I. Khomskii, Sov. Phys. Usp. {\bf 25}, 231 (1992).

\bibitem{Kugel_Khomskii_1} G. Khaliullin and V. Oudovenko, Phys. Rev. B {\bf 56}, R14243(R) (1997).

\bibitem{enhanced_pairing} P. Werner, H. U. R. Strand, S. Hoshino, Y. Murakami, and M. Eckstein, Phys. Rev. B {\bf 97}, 165119 (2018).



\bibitem{Hoshino} S. Hoshino and P. Werner, Phys. Rev. B {\bf 93}, 155161 (2016).

\bibitem{Hoshino_2} K. Steiner, S. Hoshino, Y. Nomura, and P. Werner, Phys. Rev. B {\bf 94}, 075107 (2016).


\bibitem{Aoki2014} H. Aoki, N. Tsuji, M. Eckstein, M. Kollar, T. Oka, and P. Werner, Rev. Mod. Phys. {\bf 86}, 779 (2014).

\bibitem{Eckstein2010} M. Eckstein and P. Werner, Phys. Rev. B {\bf 82}, 115115 (2010).

\bibitem{SWT} S.-S. B. Lee, J. v. Delft, and A. Weichselbaum, Phys. Rev. B {\bf 96}, 245106 (2017).


\end{thebibliography}
\end{document}

\title{Non-thermal superconductivity in photo-doped multi-orbital Hubbard systems \\ Supplemental material}
\author{Sujay Ray}
\affiliation{Department of Physics, University of Fribourg, Fribourg-1700, Switzerland}
%\email{sujay.ray@unifr.ch}
\author{Yuta Murakami}
\affiliation{Center for Emergent Matter Science, RIKEN, Wako, Saitama 351-0198, Japan}
\author{Philipp Werner}
%\email{philipp.werner@unifr.ch}
\affiliation{Department of Physics, University of Fribourg, Fribourg-1700, Switzerland}
\date{\today}
\maketitle

%\vspace{0.1in}

\section{Schrieffer-Wolff transformation}

Here we sketch the derivation of the effective Hamiltonian used in the main manuscript for the photo-doped two orbital repulsive Hubbard model. We focus on the effective Hamiltonian projected onto the triplon (triply occupied sites) and singlon (singly occupied sites) subspaces. A detailed derivation and more general analysis can be found in Ref.~\cite{SWT}. We start with the two-orbital Hubbard model with Hund coupling given in Eq. (1) in the main manuscript, which reads
\begin{eqnarray}
H &=& -t_\text{hop}\sum_{\left<ij\right>,\sigma}\sum_{\alpha=1,2} c^{\dagger}_{i,\alpha\sigma}c_{j,\alpha\sigma} + U\sum_{i}\sum_{\alpha=1,2}n_{i,\alpha\uparrow}n_{i,\alpha\downarrow}  \nonumber \\
&& \hspace{0cm} - \mu\sum_{i}\sum_{\alpha=1,2}\left(n_{i,\alpha\uparrow} + n_{i,\alpha\downarrow} \right) + (U-2J)\sum_{i,\sigma}n_{i,1\sigma}n_{i,2\Bar{\sigma}} + (U-3J)\sum_{i,\sigma}n_{i,1\sigma}n_{i,2\sigma},
\label{eq_1}
\end{eqnarray}
where the first term of Eq.~\eqref{eq_1} is the kinetic (hopping) term $H_{t_{\text{hop}}}$, the second term is the intra-orbital repulsive Hubbard interaction $H_U$, the third one the chemical potential term $H_{\mu}$, while the last two terms describe inter-orbital density-density interactions, which include the Hund coupling $J$. Our main objective is to get an effective low-energy Hamiltonian for high $U$ (strong coupling limit) and to project the effective Hamiltonian to different occupation number sectors for nearest neighbor sites $i$ and $j$. 

First, we consider the case without Hund coupling ($J=0$) and perform a generalized Schrieffer-Wolff transformation (SWT) using a rotating-frame time evolution \cite{SWT}. We define the projected operators (or Hubbard operators) from the creation and annihilation operators as
\begin{eqnarray}
c^{\dagger}_{i,\alpha\sigma} &=& \sum_{n=1}^{4}P_{in}c^{\dagger}_{i,\alpha\sigma} = \sum_{n=1}^{4} \tilde{c}^{\dagger}_{i,\alpha\sigma,n}, \\
c_{i,\alpha\sigma} &=& \sum_{n=1}^{4}c_{i,\alpha\sigma}P_{in} = \sum_{n=1}^{4} \tilde{c}_{i,\alpha\sigma,n},
\end{eqnarray}
where $P_{in}$ is the projector to the states at site $i$ with $n$ particles. Hence, $\tilde{c}^{\dagger}_{i,\alpha\sigma,n}$ creates a particle in orbital $\alpha$ and with spin $\sigma$ at site $i$, resulting in a state with $n$ particles at site $i$. Similarly, $\tilde{c}_{i,\alpha\sigma,n}$ annihilates a particle at site $i$ in a state that has $n$ particles. Alternatively, these operators can also be written as $\tilde{c}^{\dagger}_{i,\alpha\sigma,n} = c^{\dagger}_{i,\alpha\sigma}P_{in-1}$ and $\tilde{c}_{i,\alpha\sigma,n} = P_{in-1}c_{i,\alpha\sigma}$. With the introduction of these projected operators, the kinetic (hopping) term can be written as 
\begin{eqnarray}
H_{t_{\text{hop}}} = \sum_{n,n^{\prime}} H_{t_{\text{hop}},nn^{\prime}},
\label{eq_kinetic1}
\end{eqnarray}
where $H_{t_{\text{hop}},nn^{\prime}}$ is given by
\begin{eqnarray}
H_{t_{\text{hop}},nn^{\prime}} = -t_\text{hop}\sum_{\left<ij\right>} \sum_{\alpha\sigma} \tilde{c}^{\dagger}_{i,\alpha\sigma,n} \tilde{c}_{j,\alpha\sigma,n^{\prime}} = -t_\text{hop}\sum_{\left<ij\right>} \sum_{\alpha\sigma} c^{\dagger}_{i,\alpha\sigma} c_{j,\alpha\sigma} P_{in-1}P_{jn^{\prime}}.
\label{eq_kinetic2}
\end{eqnarray}
It can be seen from Eqs.~\eqref{eq_kinetic1},~\eqref{eq_kinetic2} that the kinetic term is divided into different sectors with fixed particle numbers at sites $i$ and $j$. With this notation we can proceed to perform a rotating-frame transformation generated by the term of the Hubbard Hamiltonian with the largest energy scale, $H_U$. We thus arrive at the rotating frame Hamiltonian $H_\text{rot}$ given by 
\begin{eqnarray}
H_\text{rot} =  -H_U + e^{iH_Ut}He^{-iH_Ut}.
\end{eqnarray}
The last term in the above expression can be expanded using the Baker-Campbell-Hausdorff (BCH) formula, which requires the evaluation of $[H_U,H]$. Since $H_U$ commutes with all the other terms except $H_{t_{\text{hop}}}$, the commutator can be reduced to $[H_U,H_{t_{\text{hop}}}]$, which evaluates to
\begin{eqnarray}
[H_U,H_{t_{\text{hop}}}] = \sum_{n,n^{\prime}} [H_U,H_{t_{\text{hop}},nn^{\prime}}] = \sum_{n,n^{\prime}} (n - n^{\prime})UH_{t_{\text{hop}},nn^{\prime}}.
\end{eqnarray}
We can further simplify the notation by introducing a new index $m=n-n^{\prime}$, with which the commutator can be written as 
\begin{eqnarray}
\left[H_U,H_{t_{\text{hop}}}\right] &=& \sum_{m} mU \left( \sum_{n} H_{t_{\text{hop}},n \, n-m} \right) \nonumber \\
&\equiv& \sum_{m} mUH_{t_{\text{hop}},m},
\end{eqnarray}
where $m$ can take values from $-3$ to $3$ with the constraint $4 > n-m \ge 0$ and $m \ne 0$. With these results, one can derive the following expression for $H_\text{rot}$,
\begin{eqnarray}
H_\text{rot} =  H_{\mu} + \sum_m H_{t_{\text{hop}},m}e^{imUt}.
\end{eqnarray}
This is a time periodic Hamiltonian with frequency $U$. By averaging out the high frequency dynamics, % $(>U)$, 
we obtain the low-energy effective Hamiltonian describing the nonstroboscopic dynamics, which is given by
\begin{eqnarray}
H_\text{eff} &=& H_{\mu} + H_{t_{\text{hop}},0} + H_{t_{\text{hop}}^2/U} + O(t_{hop}^3/U), \\
H_{t_{\text{hop}},0} &=& -t_\text{hop}\sum_{\left<ij\right>n}\sum_{\alpha\sigma} c^{\dagger}_{i,\alpha\sigma} c_{j,\alpha\sigma} P_{in-1}P_{jn} + (i \leftrightarrow j),\\
H_{t_{\text{hop}}^2/U} &=& \sum_{m \ne 0} \frac{H_{t_{\text{hop}},m}H_{t_{\text{hop}},-m}}{mU}.
\end{eqnarray}
By inserting the expressions for $H_{t_{\text{hop}},m}$ and and neglecting three site terms, we arrive at 
\begin{eqnarray}
H_{t_{\text{hop}}^2/U} &=& \frac{t_{\text{hop}}^2}{U} \sum_{\left<ij\right>} \sum_{n} \left( \sum_{\substack{\alpha,\alpha^{\prime} \\ \sigma,\sigma^{\prime}}} \sum_{\alpha\sigma\ne\alpha^{\prime}\sigma^{\prime}} c^{\dagger}_{i,\alpha\sigma} c^{\dagger}_{i,\alpha^{\prime}\sigma^{\prime}} c_{j,\alpha^{\prime}\sigma^{\prime}} c_{j,\alpha\sigma} \right) P_{in-2}P_{jn} \nonumber \\
&& \hspace{0cm} + \frac{t_{\text{hop}}^2}{U} \sum_{\left<ij\right>} \sum_{n, m} \frac{1}{m}\left( n_i - \sum_{\substack{\alpha,\alpha^{\prime} \\ \sigma,\sigma^{\prime}}}c^{\dagger}_{i,\alpha\sigma} c_{i,\alpha^{\prime}\sigma^{\prime}} c^{\dagger}_{j,\alpha^{\prime}\sigma^{\prime}} c_{j,\alpha\sigma} \right) P_{in}P_{jn-m-1} + (i \leftrightarrow j).
\label{eq_eff}
\end{eqnarray}

In the completely photo-doped two-orbital system, each site is either triply occupied (triplons) or singly occupied (singlons). Therefore, we project the effective Hamiltonian to the triplon-singlon subspace by choosing appropriate values for $n$ and $m$ to obtain the the projection operators $P_{i1}$ ($P_{j1}$) and $P_{i3}$ ($P_{j3}$). Then the projected Hamiltonian $H_{t_{\text{hop}}^2/U}$ can be divided into three parts -- the triplon-triplon part $H^{tt}_{t_{\text{hop}}^2/U}$ ($P_{i3}P_{j3}$), singlon-singlon part $H^{ss}_{t_{\text{hop}}^2/U}$ ($P_{i1}P_{j1}$) and singlon-triplon part $H^{st}_{t_{\text{hop}}^2/U}$ ($P_{i1}P_{j3}$). Note that the three-site terms do not contribute (within the triplon-singlon subspace) in a completely photo-doped system. The first term on the right hand side of Eq.~\eqref{eq_eff} can be written in terms of $\eta^{+}$ and $\eta^{-}$, whereas the second term can be decomposed into Kugel-Khomskii spin-orbital terms \cite{Kugel_Khomskii}. To achieve the Kugel-Khomskii decomposition we write the operator $c^{\dagger}_{i\alpha\sigma} c_{i\alpha^{\prime}\sigma^{\prime}}$ as the product $\tau s$  (where $\tau$ is the orbital-pseudospin and $s$ is the spin operator) with the following rule
\begin{eqnarray}
(\alpha,\alpha^{\prime})\hspace{0.1cm} &=& \hspace{0.1cm}(1,1) \rightarrow \frac{1}{2}+\tau^z; \hspace{0.3cm} (2,2) \rightarrow \frac{1}{2}-\tau^z; \hspace{0.3cm} (1,2) \rightarrow \tau^+; \hspace{0.3cm} (2,1) \rightarrow \tau^-,\nonumber \\
(\sigma,\sigma^{\prime})\hspace{0.1cm} &=& \hspace{0.1cm} (\uparrow,\uparrow) \rightarrow \frac{1}{2}+s^z; \hspace{0.3cm} (\downarrow,\downarrow) \rightarrow \frac{1}{2}-s^z; \hspace{0.3cm} (\uparrow,\downarrow) \rightarrow s^+; \hspace{0.3cm} (\downarrow,\uparrow) \rightarrow s^-.
\end{eqnarray}
This leads us to the expression for the effective Hamiltonian for $J=0$ which is shown in Eqs. (6), (7) of the main manuscript:
\begin{eqnarray}
\label{eq_7}
H_{t_{\text{hop}}^2/U}^{st} &=& -\frac{4t_{hop}^2}{U} \sum_{\left<ij\right>,\alpha} \left( \eta_{i\alpha}^{+}\eta_{j\alpha}^{-} + \eta_{i\alpha}^{-}\eta_{j\alpha}^{+} \right) - \frac{4t_{hop}^2}{3U} \sum_{\left<ij\right>} \left( \frac{1}{2} + 2\boldsymbol{s_i} . \boldsymbol{s_j} \right) \left( \frac{1}{2} + 2\boldsymbol{\tau_i} . \boldsymbol{\tau_j} \right), \\
%H_{t^2/U}^{ss} &=& H_{t^2/U}^{tt} = 
H_{t_{\text{hop}}^2/U}^{ss(tt)} &=& 
 \frac{2t_{hop}^2}{U} \sum_{\left<ij\right>} \left( \frac{1}{2} + 2\boldsymbol{s_i} . \boldsymbol{s_j} \right) \left( \frac{1}{2} + 2\boldsymbol{\tau_i} . \boldsymbol{\tau_j} \right).
\label{eq_8}
\end{eqnarray}

\begin{figure}[t]\centering
\includegraphics[width=0.98\textwidth]{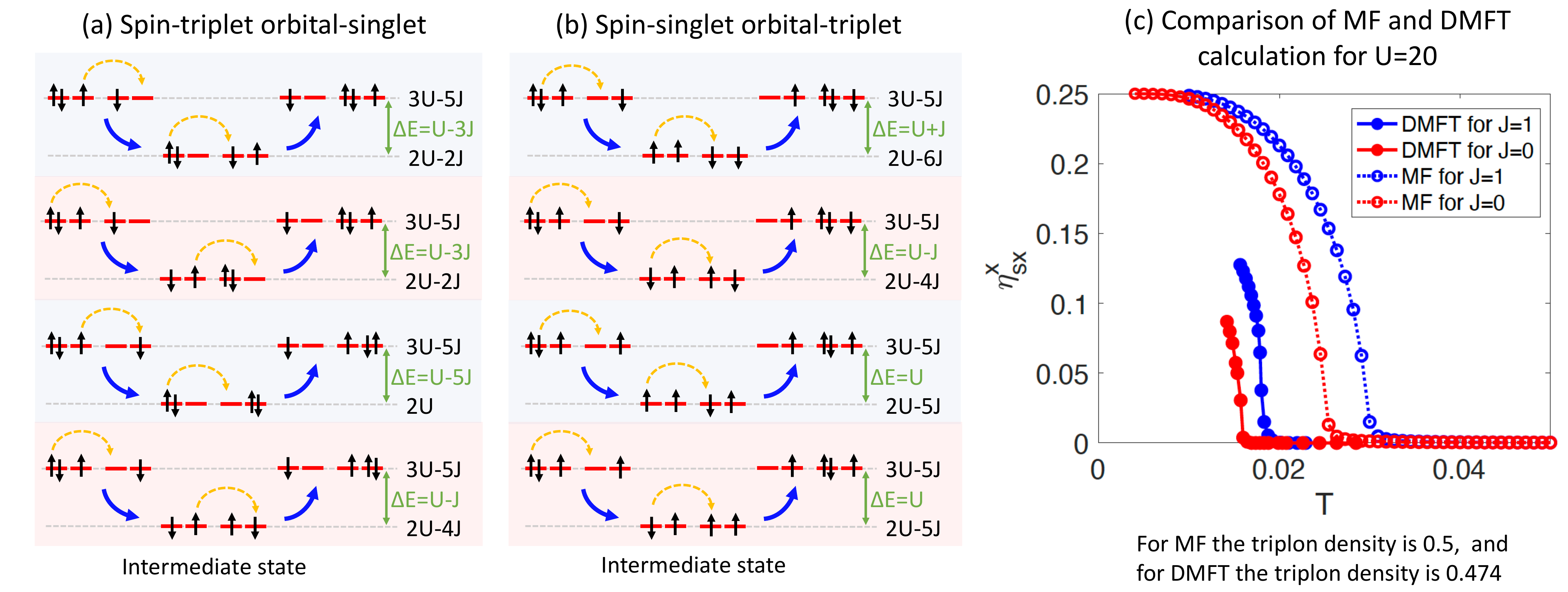}
\caption{Various second order hopping processes contributing to $\eta_{sx}$ and $\eta_{ox}$ paring. (a) $c^{\dagger}_{1\uparrow}c^{\dagger}_{2\uparrow}$ spin-triplet hopping, and (b) $c^{\dagger}_{1\uparrow}c^{\dagger}_{1\downarrow}$ spin-singlet hopping. (c) MF and DMFT $\eta^{x}_{sx}$ superconducting order for $U=20$ and for  Hund coupling $J=0$ and $1$. Both MF (for triplon density 0.474) and DMFT (for triplon density 0.5) show an enhancement of the $\eta^{x}_{sx}$ order upon inclusion of a Hund coupling $J>0$. 
}
\label{fig_SM1}
\end{figure}

The $J\ne 0$ case can be analyzed by noting that in Eq.~\eqref{eq_7} the $\eta$ terms come from second order hopping processes, where the energy difference between the initial state and the intermediate state depends on $J$. In the absence of Hund coupling ($J=0$), all the $\eta$ superconducting pairings are degenerate, since all the intermediate states differ by the same local energy $\Delta E=U$ from the initial and final state. The inclusion of Hund coupling lifts this degeneracy, as the intermediate doublon states are no longer degenerate. As shown in Fig.~\ref{fig_SM1}(a), for $J>0$ the spin-triplet orbital-singlet $\eta_{sx}$ pairing has intermediate doublon states which differ by the local energy $\Delta E= U-2J, U-3J, U-5J$ from the initial state (consisting of a singlon-triplon pair with local energy $3U-5J$). In Fig.~\ref{fig_SM1}(a), we only show the intermediate state for the hopping process $c^{\dagger}_{1\uparrow}c^{\dagger}_{2\uparrow}$. It can be easily seen that all the other processes involved in $\eta_{sx}$ pairing have the same energy configurations.

Taking into account all the intermediate states in second order perturbation theory, the effective $\Tilde{U}_{sx}$ for spin-triplet $\eta_{sx}$ pairing can be calculated as
\begin{eqnarray}
\frac{1}{\Tilde{U}_{sx}}=\frac{1}{4}\left( \frac{1}{U-J} + \frac{2}{U-3J} + \frac{1}{U-5J} \right).
\label{eq_9}
\end{eqnarray}
On the other hand, orbital-triplet spin-singlet $\eta_{ox}$ pairing has intermediate doublon states with local energy differences $\Delta E= U-J, U, U+J$ (see Fig.~\ref{fig_SM1}(b)) and thus the effective $\Tilde{U}_{ox}$ is given by
\begin{eqnarray}
\frac{1}{\Tilde{U}_{ox}}=\frac{1}{4}\left( \frac{1}{U-J} + \frac{2}{U} + \frac{1}{U+J} \right).
\label{eq_10}
\end{eqnarray}
It follows from these expressions that $J>0$ favors spin-triplet $\eta_{sx}$ pairing, because the lower $\tilde U$ results in a larger exchange coupling. Similarly, $J<0$ favors orbital-triplet $\eta_{ox}$ pairing.

\begin{figure}[t]\centering
\includegraphics[width=0.75\textwidth]{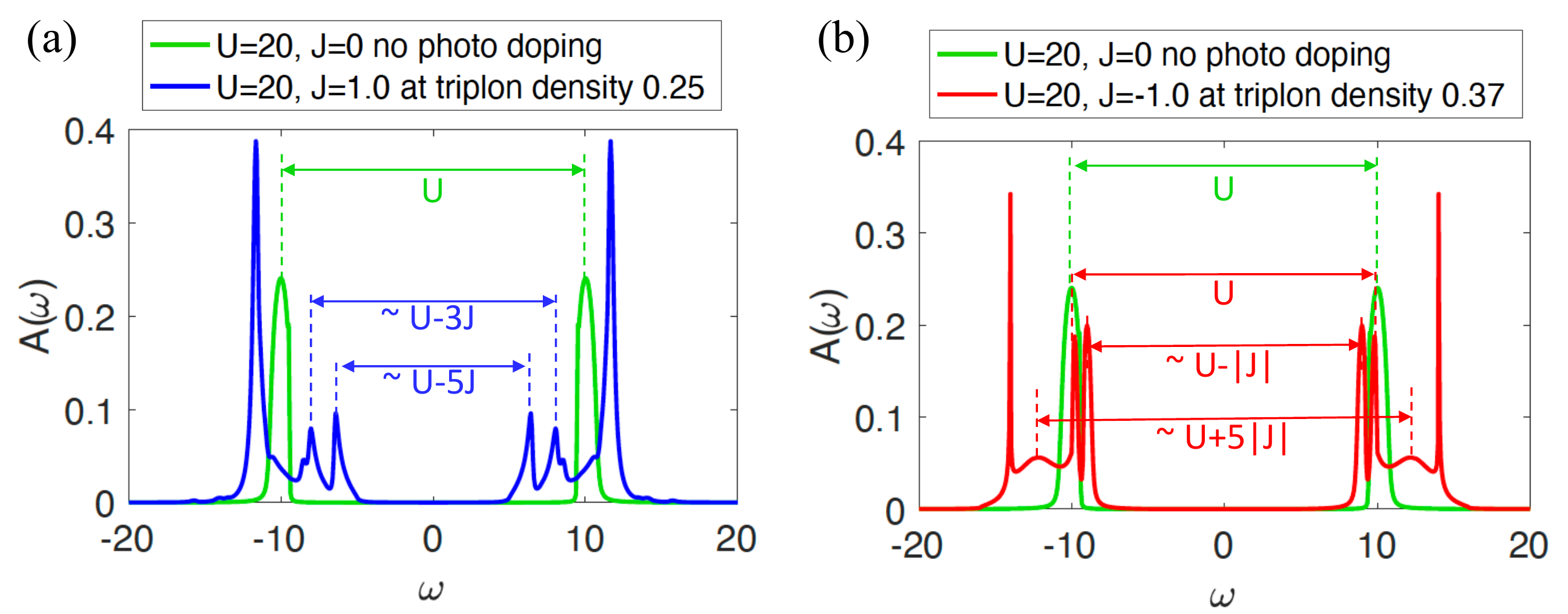}
\caption{Non-equilibrium spectra for (a) $U=20, J=1$ and (b) $U=20$, $J=-1$. Both spectra are obtained for a reduced $t_\text{hop}=0.3$ and are compared with the undoped equilibrium spectrum for $U=20, J=0$ (green). For $J>0$ (panel (a)), additional peaks appear with an energy separation of approximately $U-5J$ and $U-3J$, resulting in a smaller gap. In panel (b), for $J<0$, a prominent feature with an energy separation of approximately $U+5|J|$ appears, resulting in an overall larger separation of the Hubbard bands. 
}
\label{fig_SM2}
\end{figure}

\section{Non-equilibrium spectral function}

The non-equilibrium single-particle spectra of multi-orbital systems with Hund coupling can be very different from the equilibrium counterpart. In equilibrium, the half filled system without Hund coupling ($J=0$) has a band gap of $U$ between the lower and upper Hubbard band. Near the atomic limit ($U\gg t_\text{hop}$), for $J>0$, the two dominant Hubbard subbbands are separated by $U+J$, while for $J<0$ the band gap is $U+5|J|$. In non-equilibrium photo-doped systems, the triplon sites have a local energy $3U-5J$ and the removal of an electron from a triplon site can produce doublons with interaction energy $U$, $U-2J$ or $U-3J$. Similarly, taking into account the chemical potential term, a singlon can gain local energy by the addition of an electron. Thus in the non-equilibrium system, the splitting between the Hubbard bands, estimated as $E^{N=3}_\text{loc}+E^{N=1}_\text{loc}-2E^{N=2}_\text{loc}$, is expected to cover the energy range from approximately $U-5J$ to $U+J$.

Consistent with this expectation, in Fig.~1(b),(c) of the main manuscript, we see a smaller separation between the Hubbard bands for $J=1$ than for $J=-1$. Here, the substructures associated with different doublon states are broadened due to the kinetic term. By reducing the hopping $t_\text{hop}$, the corresponding peaks can be made sharp. In Fig.~\ref{fig_SM2}(a), we plot the spectra for $U=20, J=1, t_\text{hop}=0.3$. One can now clearly identify the features produced by the different doublon states upon removal of an electron from a triplon (upper Hubbard band) or addition of an electron to a singlon (lower Hubbard band). As a result of these peaks, the gap shrinks, compared to $J=0$. Similarly, Fig.~\ref{fig_SM2}(b) shows that for $J=-1$, the additional peaks induced by photo-doping lead to a broadening and enhanced splitting of the Hubbard bands.

\section{Mean field decoupling}

%\begin{figure}[t]\centering
%\includegraphics[width=0.45\textwidth]{fig_SM3.pdf}
%\caption{Spin-triplet orbital-singlet superconducting order $\eta^{x}_{sx}$ as a function of triplon density for different seed fields $P_\text{seed}$ for $U=20$, $J=1$ and $\beta_\text{eff}=60$. In the superconducting state (triplon density above 0.463) the values of the order parameter become essentially independent of the seed field, which indicates a spontaneous symmetry breaking. 
%}
%\label{fig_SM2}
%\end{figure}

We perform the same-site mean field (MF) decoupling on the effective model given by Eqs.~\eqref{eq_7},~\eqref{eq_8}, which  (neglecting a constant term) reads
\begin{eqnarray}
\label{eq_9}
&&H_{t_{\text{hop}}^2/U}^{ts,\text{MF}} = -\frac{4t_{\text{hop}}^2}{U} \sum_{\left<ij\right>} \left[ \sum_{\alpha}\left( \left<\eta_{i\alpha}^{+}\right> \eta_{j\alpha}^{-} + \left<\eta_{i\alpha}^{-}\right> \eta_{j\alpha}^{+} \right) - \frac{1}{3} \left( \left<\boldsymbol{s_i}\right> . \boldsymbol{s_j} + \left<\boldsymbol{\tau_i}\right> . \boldsymbol{\tau_j} + 4\left<\boldsymbol{s_i} \boldsymbol{\tau_i}\right>  \boldsymbol{s_j} \boldsymbol{\tau_j} \right) \right] + \left( i \rightarrow j \right), \nonumber \\
%H_{t^2/U}^{ss} &=& H_{t^2/U}^{tt} = 
&&H_{t_{\text{hop}}^2/U}^{ss(tt),\text{MF}} = 
 \frac{2t_{\text{hop}}^2}{U} \sum_{\left<ij\right>} \left[ \left<\boldsymbol{s_i}\right> . \boldsymbol{s_j} + \left<\boldsymbol{\tau_i}\right> . \boldsymbol{\tau_j} + 4\left<\boldsymbol{s_i} \boldsymbol{\tau_i}\right>  \boldsymbol{s_j} \boldsymbol{\tau_j} \right] + \left( i \rightarrow j \right),
\label{eq_10}
\end{eqnarray}
where the spin-orbital composite order $\boldsymbol{s}_i\boldsymbol{\tau}_i$ can be defined as a tensor $(s_i\tau_i)^{\mu\nu}=\sum_{\alpha\sigma}\sum_{\alpha^{\prime}\sigma^{\prime}}c^{\dagger}_{i,\alpha\sigma}\sigma^{\mu}_{\sigma\sigma^{\prime}}\tau^{\nu}_{\alpha\alpha^{\prime}}c_{i,\alpha^{\prime}\sigma^{\prime}}$. This MF Hamiltonian can be solved self-consistently for $\eta$ order. We calculate the MF $\eta$ order for the extremely photo-doped (triplon density 0.5) system. As mentioned earlier, in this limit, each site is either a triplon or a singlon. So, all the operators used in the MF Hamiltonian can be written in the triplon-singlon basis. First, we note that there are four triplons and four singlons denoted by
\begin{eqnarray}
&&|t_{1\uparrow}\rangle_i=t_{i,1\uparrow}^{\dagger}|0\rangle_i = c^{\dagger}_{i,1\uparrow}c^{\dagger}_{i,2\uparrow}c^{\dagger}_{i,2\downarrow}|0\rangle_i, \hspace{1.5cm} |s_{1\uparrow}\rangle_i=s_{i,1\uparrow}^{\dagger}|0\rangle_i = c^{\dagger}_{i,1\uparrow}|0\rangle_i, \nonumber \\
&&|t_{1\downarrow}\rangle_i=t_{i,1\downarrow}^{\dagger}|0\rangle_i =c^{\dagger}_{i,1\downarrow}c^{\dagger}_{i,2\uparrow}c^{\dagger}_{i,2\downarrow}|0\rangle_i, \hspace{1.5cm} |s_{1\downarrow}\rangle_i=s_{i,1\downarrow}^{\dagger}|0\rangle_i = c^{\dagger}_{i,1\downarrow}|0\rangle_i, \nonumber \\
&&|t_{2\uparrow}\rangle_i=t_{i,2\uparrow}^{\dagger}|0\rangle_i =c^{\dagger}_{i,1\uparrow}c^{\dagger}_{i,1\downarrow}c^{\dagger}_{i,2\uparrow}|0\rangle_i, \hspace{1.5cm} |s_{2\uparrow}\rangle_i=s_{i,2\uparrow}^{\dagger}|0\rangle_i = c^{\dagger}_{i,2\uparrow}|0\rangle_i, \nonumber \\
&&|t_{2\downarrow}\rangle_i=t_{i,2\downarrow}^{\dagger}|0\rangle_i =c^{\dagger}_{i,1\downarrow}c^{\dagger}_{i,1\downarrow}c^{\dagger}_{i,2\downarrow}|0\rangle_i, \hspace{1.5cm} |s_{2\downarrow}\rangle_i=s_{i,2\downarrow}^{\dagger}|0\rangle_i = c^{\dagger}_{i,2\downarrow}|0\rangle_i,
\end{eqnarray}
where $|t_{\alpha\sigma}\rangle_i$ denotes the triplon state, $|s_{\alpha\sigma}\rangle_i$ denotes the singlon state and $|0\rangle_i$ is the vacuum state at site $i$. Now, we can define the triplon basis as $\mathcal{T}^{\dagger}=(t_{1\uparrow}^{\dagger}, \hspace{0.2cm} t_{1\downarrow}^{\dagger}, \hspace{0.2cm} t_{2\uparrow}^{\dagger}, \hspace{0.2cm} t_{2\downarrow}^{\dagger})$ and the singlon basis as $\mathcal{S}^{\dagger} = (s_{1\uparrow}^{\dagger}, \hspace{0.2cm} s_{1\downarrow}^{\dagger}, \hspace{0.2cm} s_{2\uparrow}^{\dagger}, \hspace{0.2cm} s_{2\downarrow}^{\dagger})$. The triplon-singlon basis is then written as $\psi^{\dagger}=(\mathcal{T}^{\dagger}, \hspace{0.3cm} \mathcal{S}^{\dagger})$ which we use to write the MF operators. For example, the $\eta$ operator takes a singlon and converts it into a triplon and vice versa. So, $\eta$ operators correspond to off-diagonal block matrices. On the other hand, all the spin and orbital operators are block diagonal in this basis.

\begin{figure}[t]\centering
\includegraphics[width=0.97\textwidth]{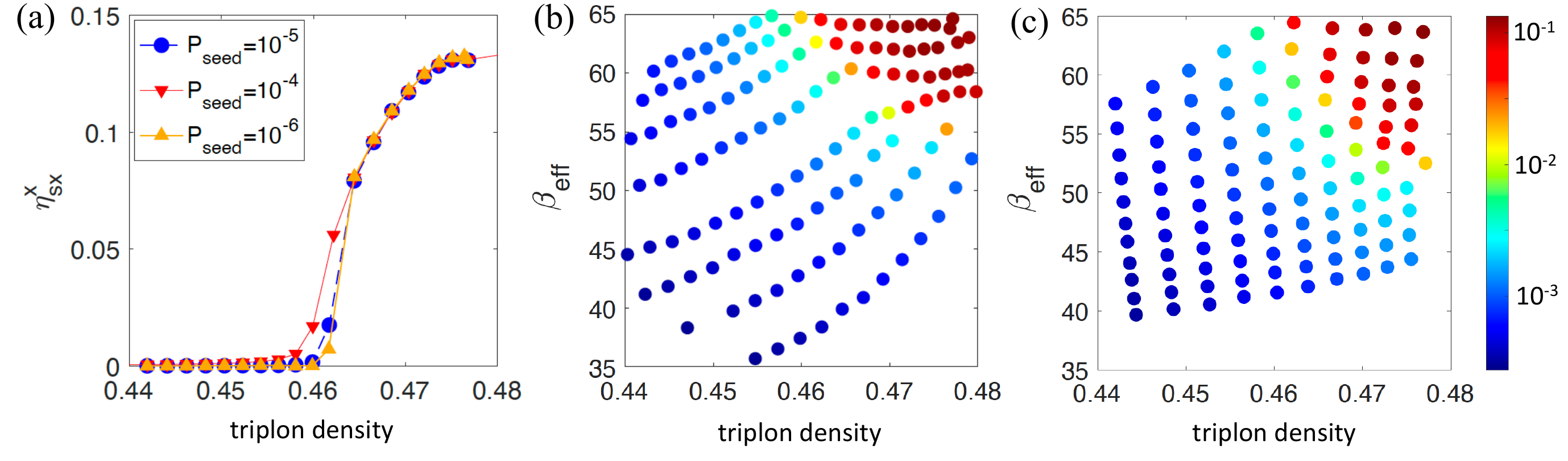}
\caption{(a) Spin-triplet orbital-singlet superconducting order $\eta^{x}_{sx}$ as a function of triplon density for different seed fields $P_\text{seed}$ for $U=20$, $J=1$, $\Gamma=0.02$ and $\beta_\text{eff}=60$. In the superconducting state (triplon density above 0.463) the values of the order parameter become essentially independent of the seed field, which indicates a spontaneous symmetry breaking. Panels (b) and (c) show the phase diagrams in the $\beta_\text{eff}$ and triplon density space for $U=20$ and $J=1$. The effective temperature $\beta_\text{eff}$ is controlled by tuning two different parameters -- the coupling to the Fermion bath ($\Gamma=0.01$ to $0.03$) in (b) and the temperature of the Fermion bath (keeping the coupling constant at $\Gamma=0.02$) in (c). The color map indicates the order parameter.
}
\label{fig_SM3}
\end{figure}

Figure~\ref{fig_SM1}(c) shows the MF calculation for the order parameter $\eta^{x}_{sx}=$ Re[$\eta^{+}_{sx}$] as a function of temperature $T$ for $U=20$ and triplon density 0.5. In the absence of Hund coupling ($J=0$), the MF SC transition temperature is $T^\text{MF}_c\sim 0.026$ (red dashed line), whereas the DMFT calculation shows a SC transition temperature $T^\text{DMFT}_c\sim 0.018$ (red line) for triplon density 0.474. This over-estimation may be due to the fact that we cannot reach the extreme doping (triplon density 0.5) limit in DMFT and also because the MF calculation is based on an effective model which is valid in the $U\gg t$ limit. The MF treatment also neglects temporal fluctuations and the coupling to Fermion baths, which may contribute to the comparatively lower $T_{c}$ in the DMFT calculations. However, consistent with the DMFT results, we observe that the inclusion of a $J>0$ Hund coupling term favors the spin-triplet orbital-singlet $\eta$ order, as shown in Fig.~\ref{fig_SM1}(c) where $T^\text{MF}_c$  is enhanced (blue dashed line). The DMFT calculation shows the same trend (solid blue line).

\section{DMFT calculations}

As mentioned in the main text, in the DMFT calculations the Green's function $G(t,t^{\prime})$ and the hybridization function $\Delta(t,t^{\prime})$ are expressed in the Nambu basis. In order to measure the spin-triplet superconducting order, we introduce the spinor $\psi^{\dagger}=(c_{1\uparrow}^{\dagger}, \hspace{0.2cm} c_{2\uparrow}, \hspace{0.2cm} c_{1\downarrow}^{\dagger}, \hspace{0.2cm} c_{2\downarrow})$ so that the hybridization function and the Green's function can be written in this basis as $4\times4$ matrices
\begin{eqnarray}
\Delta(t,t^{\prime}) &=& \left[
\begin{array}{ c c c c }
\Delta_{1\uparrow 1\uparrow}^{c^{\dagger}c} & \Delta_{1\uparrow 2\uparrow}^{c^{\dagger}c^{\dagger}} & 0 & \Delta_{1\uparrow 2\downarrow}^{c^{\dagger}c^{\dagger}}\\
\Delta_{2\uparrow 1\uparrow}^{cc} & \Delta_{2\uparrow 2\uparrow}^{cc^{\dagger}} & \Delta_{2\uparrow 1\downarrow}^{cc} & 0 \\
 0 & \Delta_{1\downarrow 2\uparrow}^{c^{\dagger}c^{\dagger}} & \Delta_{1\downarrow 1\downarrow}^{c^{\dagger}c} & \Delta_{1\downarrow 2\downarrow}^{c^{\dagger}c^{\dagger}}\\
 \Delta_{2\downarrow 1\uparrow}^{cc} & 0 & \Delta_{2\downarrow 1\downarrow}^{cc} & \Delta_{2\downarrow 2\downarrow}^{cc^{\dagger}}\\
\end{array} \right], \ \
G(t,t^{\prime})=\left[
\begin{array}{ c c c c }
G_{1\uparrow 1\uparrow}^{cc^{\dagger}} & G_{1\uparrow 2\uparrow}^{cc} & 0 & G_{1\uparrow 2\downarrow}^{cc} \\
G_{2\uparrow 1\uparrow}^{c^{\dagger}c^{\dagger}} & G_{2\uparrow 2\uparrow}^{c^{\dagger}c} & G_{2\uparrow 1\downarrow}^{c^{\dagger}c^{\dagger}} & 0 \\
 0 & G_{1\downarrow 2\uparrow}^{cc} & G_{1\downarrow 1\downarrow}^{cc^{\dagger}} & G_{1\downarrow 2\downarrow}^{cc}\\
 G_{2\downarrow 1\uparrow}^{c^{\dagger}c^{\dagger}} & 0 & G_{2\downarrow 1\downarrow}^{c^{\dagger}c^{\dagger}} & G_{2\downarrow 2\downarrow}^{c^{\dagger}c}\\
\end{array} \right]. \nonumber
%&=&\left[
%\begin{array}{ c c }
%q(x_2^2-2x_1^2)  & -q(x_2y_2+2x_1y_1)\\
%-q(x_2y_2+2x_1y_1) & q(y_2^2-2y_1^2)
%\end{array} \right]
\end{eqnarray}
The DMFT self consistency equations in this Nambu basis can be written as $\Delta(t,t^{\prime})=t_{0}^2\gamma G(t,t^{\prime})\gamma$ . For uniform superconducting order $\gamma$ is a diagonal matrix given by $\gamma=$ diag( $1, \hspace{0.2cm} -1, \hspace{0.2cm} 1, \hspace{0.2cm} -1$ ) and for staggered $\eta$ superconducting order $\gamma$ is a unity matrix $\gamma=$ diag( $1, \hspace{0.2cm} 1, \hspace{0.2cm} 1, \hspace{0.2cm} 1$ ). Please note that to measure the spin-triplet order we assume there is no intra-orbital superconducting order like $c^{\dagger}_{\alpha\uparrow}c^{\dagger}_{\alpha\downarrow}$. In a similar manner, for measuring the orbital-triplet order, we assume no inter-orbital pairing and write the hybridization function and Green's function in the spinor basis $\psi^{\dagger}=(c_{1\uparrow}^{\dagger}, \hspace{0.2cm} c_{1\downarrow}, \hspace{0.2cm} c_{2\uparrow}^{\dagger}, \hspace{0.2cm} c_{2\downarrow})$. For the derivation of these DMFT equations, one can follow Ref.~\cite{enhanced_pairing}.

To measure a specific superconducting order, we apply a small seed field $P_\text{seed}$ and couple it to the superconducting order and keep it on during self-consistent DMFT calculation. In a symmetry broken state, except close to the phase transition point, the order parameter should not strongly depend on the seed field. We demonstrate this in Fig.~\ref{fig_SM3}(a), where seed fields of different orders of magnitude ($10^{-4},10^{-5},10^{-6}$) were applied to measure $\eta^{x}_{sx}$ for $U=20$ and $J=1$. The value of the order parameter $\eta^{x}_{sx}$ is essentially independent of the seed field above triplon density 0.463, which indicates the existence of a symmetry broken state. We have also checked that the value of the order parameter remains fixed even if we switch off the seed field after a certain number of DMFT iterations. For triplon densities below 0.463, the order parameter is reduced if the seed field is reduced, suggestive of a normal state. Our data indicate a discontinuous transition from the normal to the superconducting phase at the given effective temperature ($T_\text{eff}=1/60$).

The non-equilibrium steady state corresponds to a partially thermalized photo-doped state, where the electrons are thermalized within each Hubbard band. Such a state should be described by a well-defined effective temperature which we can find by comparing the distribution function of the electrons in each Hubbard band to an appropriately shifted Fermi distribution function. In order to demonstrate that the non-equilibrium state is indeed described by an effective temperature $T_\text{eff}=1/\beta_\text{eff}$, we cool the system by varying different parameters. Our calculation shows that the steady state is described by the same order parameter at a fixed $\beta_\text{eff}$ and it does not depend on how this temperature is achieved. Figure~\ref{fig_SM3}(b),(c) shows the phase diagram of photo-doped systems with $U=20$ and $J=1$, obtained by tuning the coupling to the Fermion bath (Fig.~\ref{fig_SM3}(b)) and by tuning the temperature of the Fermion bath (Fig.~\ref{fig_SM3}(c)). It can be seen that the phase diagram is essentially independent of the tuning parameter used to cool down the system.

\begin{figure}[t]\centering
\includegraphics[width=0.96\textwidth]{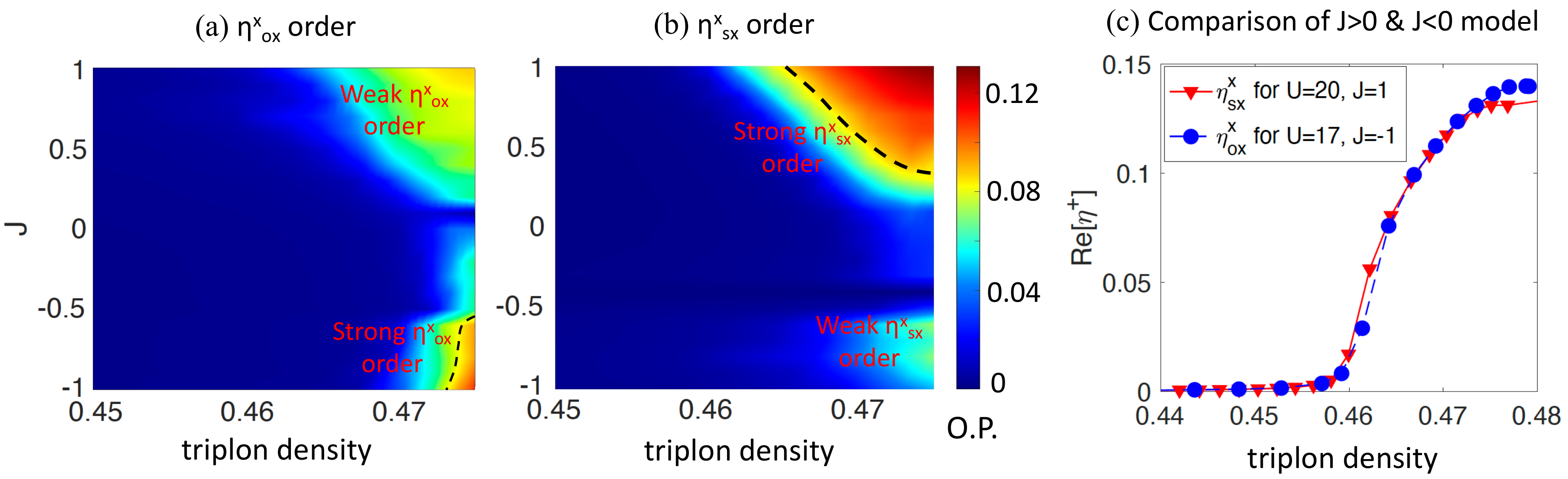}
\caption{Nonquilibrium phase diagrams for $\eta$ superconductivity in the space of triplon density and Hund coupling. The orbital-triplet order $\eta^{x}_{ox}$ is plotted in (a) and spin-triplet order $\eta^{x}_{sx}$ is plotted in (b), with the color map indicating the order parameter (O.P.). $\eta^{x}_{sx}$ dominates the $J>0$ region, while $\eta^{x}_{ox}$ dominates the $J<0$ region. The qualitative equivalence of the $J>0$ and $J<0$ models is shown in panel (c) by plotting $\eta^{x}_{sx}$ for $U=20, J=1$ and $\eta^{x}_{ox}$ for $\Tilde{U}=U+3|J|=20, J=-1$ as a function of the triplon density at $\beta_\text{eff}=60$.
}
\label{fig_SM4}
\end{figure}

Finally, we show  in Fig.~\ref{fig_SM4} the phase diagram for a photo-doped system with $U=20$ and $\beta_\text{eff}=60$ in the space spanned by the Hund coupling $J$ and triplon density. Our calculation shows that indeed $\eta$ pairing is favored as we increase the Hund coupling $J$, which is qualitatively consistent with the MF result. In Fig.~\ref{fig_SM4} we plot both the spin-singlet orbital-triplet order $\eta^{x}_{ox}$ (Fig.~\ref{fig_SM4}(a)) and the spin-triplet orbital-singlet order $\eta^{x}_{sx}$ (Fig.~\ref{fig_SM4}(b)). We observe that both superconducting orders are stabilized in the presence of a non-zero $J$. But the $\eta^{x}_{sx}$ order has higher values for $J>0$. This suggests that for positive values of $J$, the dominant superconducting order is given by $\eta_{sx}$, whereas for negative $J$ the orbital-triplet order $\eta_{ox}$ is the dominant order. The resulting phase diagram is consistent with the fact that the $J<0$ model can be mapped qualitatively to the $J>0$ model by flipping the spin and orbital degrees of freedom as 
$c_{i,1\downarrow} \rightarrow c_{i,2\uparrow}$ and $c_{i,2\uparrow} \rightarrow c_{i,1\downarrow}$ \cite{Hoshino_2}. After this transformation the density-density interaction terms of the Hamiltonian become
\begin{eqnarray}
H_U &=& \frac{\Tilde{U}}{2}\sum_{i}\sum_{\alpha=1,2}\sum_{\sigma\sigma^{\prime}}n_{i,\alpha\sigma}n_{i,\alpha\sigma^{\prime}}, \\
H_{J}&=& -J\sum_{i,\sigma}n_{i,1\sigma}n_{i,2\Bar{\sigma}} - 3J\sum_{i,\sigma}n_{i,1\sigma}n_{i,2\sigma} = |J|\sum_{i,\sigma}n_{i,1\sigma}n_{i,2\Bar{\sigma}} + 3|J|\sum_{i,\sigma}n_{i,1\sigma}n_{i,2\sigma}.
\label{eq_tr}
\end{eqnarray}
Defining $\Tilde{U}=U-3J=U+3|J|$, the intra-orbital repulsion for negative $J$ becomes $\Tilde{U}=U+3|J|$, the inter-orbital repulsion for same spins $\Tilde{U}-3|J|=U$ and the inter-orbital repulsion for opposite spins $\Tilde{U}-|J|=U+2|J|$. Moreover, since the spin and orbital degrees of freedom are exchanged in this transformation, the superconducting order parameter changes as
\begin{eqnarray}
(c^{\dagger}_{1\uparrow}c^{\dagger}_{2\uparrow}-c^{\dagger}_{1\downarrow}c^{\dagger}_{2\downarrow}) \rightarrow (c^{\dagger}_{1\uparrow}c^{\dagger}_{1\downarrow}-c^{\dagger}_{2\uparrow}c^{\dagger}_{2\downarrow}).
\end{eqnarray}
Thus we expect an orbital-triplet $\eta$ order for $J<0$. The higher effective $\Tilde{U}$ explains the smaller superconducting region in the phase diagram for negative $J$. This qualitative equivalence of the $J>0$ and $J<0$ models is further verified in Fig.~\ref{fig_SM4}(c), where we compare the $\eta_{sx}$ order for the $U=20$, $J=1$ model to the $\eta_{ox}$ order for the $U=17$, $J=-1$ ($\Tilde{U}=20$) model as a function of triplon density at $\beta_\text{eff}=60$. Both curves (red and blue) almost fall on top of each other, which demonstrates the validity of the mapping.